\documentclass[aps,prl,twocolumn,showpacs,groupedaddress]{revtex4-1}
\usepackage{graphicx}  
\usepackage{dcolumn}   
\usepackage{bm}        
\usepackage{amssymb}   
\usepackage{verbatim}
\usepackage{amsmath}
\usepackage{amsfonts}

\def\lsim{\mathrel{\rlap{\lower4pt\hbox{\hskip1pt$\sim$}}
    \raise1pt\hbox{$<$}}}                
\def\gsim{\mathrel{\rlap{\lower4pt\hbox{\hskip1pt$\sim$}}
    \raise1pt\hbox{$>$}}}                

\begin{document}

\title{SLUG Microwave Amplifier: Theory}
\author{G. J. Ribeill}
\author{D. Hover}
\author{Y.-F. Chen}
\author{S. Zhu}
\author{R. McDermott}
\email[Electronic address: ]{rfmcdermott@wisc.edu}

\affiliation{Department of Physics, University of
Wisconsin, Madison, Wisconsin 53706, USA}
\date{\today}

\begin{abstract}
We describe a novel scheme for low-noise phase-insensitive linear amplification at microwave frequencies based on the Superconducting Low-inductance Undulatory Galvanometer (SLUG). Direct integration of the junction equations of motion provides access to the full scattering matrix of the SLUG. We discuss the optimization of SLUG amplifiers and calculate amplifier gain and noise temperature in both the thermal and quantum regimes. Loading of the SLUG element by the finite input admittance is taken into account, and strategies for decoupling the SLUG from the higher-order modes of the input circuit are discussed. The microwave SLUG amplifier is expected to achieve noise performance approaching the standard quantum limit in the frequency range from 5-10~GHz, with gain around 15~dB for a single-stage device and instantaneous bandwidths of order 1~GHz.
\end{abstract}

\maketitle

\section{I. Introduction}

The rapid development of superconducting quantum electronics has motivated a search for near quantum-limited microwave amplifiers for the low-noise readout of qubits and linear cavity resonators. It was long ago recognized that the dc Superconducting QUantum Interference Device (dc SQUID) can achieve noise performance approaching the fundamental quantum limit imposed on phase-insensitive linear amplifiers: namely, the amplifier must add at least half a quantum of noise to the signal it amplifies \cite{Koch80}. Yet while the SQUID is in principle capable of amplifying signals at frequencies approaching the Josephson frequency (typically of order tens of GHz), it remains challenging to embed the SQUID in a 50~$\Omega$ environment and to provide for efficient coupling of a microwave signal to the device. Recently, it was shown that near quantum-limited performance can be achieved with a microstrip SQUID amplifier, where the input coil is configured as a microstrip resonator with the SQUID washer acting as a groundplane \cite{Mueck98}. The noise temperature of a microstrip SQUID amplifier cooled to millikelvin temperatures has been measured to be 47 $\pm$ 10 mK and 48 $\pm$ 5 mK at frequencies of 519 MHz and 612 MHz, respectively, more than an order of magnitude lower than the best semiconductor amplifiers available and within a factor of 2 of the quantum limit \cite{Mueck01, Kinion11}. However, efforts to extend the operating frequencies of these amplifiers into the gigahertz range are hampered by the fact that reduction of the length of the input resonator is coupled to reduction of the mutual inductance between the input coil and the SQUID \cite{Mueck03}. Alternative approaches have included the integration of a high-gain SQUID gradiometer into a coplanar waveguide resonator at a current antinode \cite{Spietz08, Spietz09}.

The current study was motivated by the development of a new device configuration that enables the efficient coupling of a GHz-frequency signal to a low-inductance, high gain SQUID that should achieve noise performance approaching the standard quantum limit. The gain element is more properly termed a SLUG (Superconducting Low-inductance Undulatory Galvanometer), as the signal is not coupled to the device inductively, but rather injected directly into the device loop as a current \cite{Clarke66}. The low-inductance design is straightforward to model at microwave frequencies, and the SLUG is readily incorporated into a microstrip line in such a way that the modes of the SLUG element and the input resonator remain cleanly resolved, greatly simplifying analysis of the circuit. In what follows we present a comprehensive theoretical study of the gain and noise performance of the SLUG microwave amplifier. Our goals are to clearly spell out the design tradeoffs, to outline a clear path to device optimization, and to identify the fundamental limits to performance.

As we shall see, the scattering parameters of the SLUG are very similar to those of the more familiar symmetric dc SQUID, apart from a trivial shift in flux bias that arises from the asymmetric division of bias current between the two arms of the SLUG. However, while it is straightforward to fabricate a low-inductance ($\sim$~10~pH) SLUG and to embed the device in a 50~$\Omega$ environment, it is challenging to engineer a clean, purely inductive coupling to a conventional dc SQUID at microwave frequencies. For this reason we have chosen to focus our discussion of microwave amplifiers on the SLUG geometry. In this manuscript we will not consider phase-sensitive amplifiers based on parametrically modulated Josephson junctions operated in the supercurrent state \cite{Yurke88, Yurke89}. There has been significant recent development of low-noise Josephson parametric amplifiers \cite{Castellanos07, Yamamoto08, Hatridge11}, including such milestones as squeezing of vacuum noise \cite{Castellanos08} and observation of quantum jumps in a superconducting qubit \cite{Vijay11}. Because these amplifiers squeeze the input state, they can achieve added noise numbers for one field quadrature that are below the standard quantum limit \cite{Caves82, Clerk10}. Moreover, these devices operate with negligible dissipation, circumventing practical problems associated with hot-electron effects that are intrinsic to devices that operate in the finite-voltage regime.
In related work, there have been efforts to develop low-noise phase-insensitive amplifiers based on parametrically modulated junctions in a ring modulator configuration \cite{Bergeal10}. Broadly speaking, advantages of the Josephson parametric amplifiers include unsurpassed noise performance and ease of fabrication, while potential disadvantages relative to SQUID-based dissipative amplifiers include modest gain-bandwidth product, limited dynamic range, and increased complexity of operation. Ultimately we suspect that there is a place in the superconducting quantum optician's toolbox for both ultralow noise phase-sensitive parametric amplifiers and robust, broadband phase-insensitive amplifiers operating near the standard quantum limit.

This paper is organized as follows. In Section II we introduce the circuit models of the symmetric dc SQUID and the SLUG. In Section III we calculate the dc characteristics of the devices. In Section IV we evaluate SLUG scattering parameters, and examine the maximum achievable gain over the range of device parameters. Sections V and VI present an analysis of noise properties in the thermal and quantum regimes, respectively. In Section VII we describe the design and performance of practical SLUG amplifiers for GHz frequency operation, and in Section VIII we discuss amplifier dynamic range. In Section IX we describe the effect of the finite admittance of the input circuit on device characteristics, gain, and noise, and in Section X we discuss hot-electron effects. In Section XI we present our concluding remarks.

\section{II. Device Model}

\begin{figure}[t]
\begin{center}
\includegraphics[width=.45\textwidth]{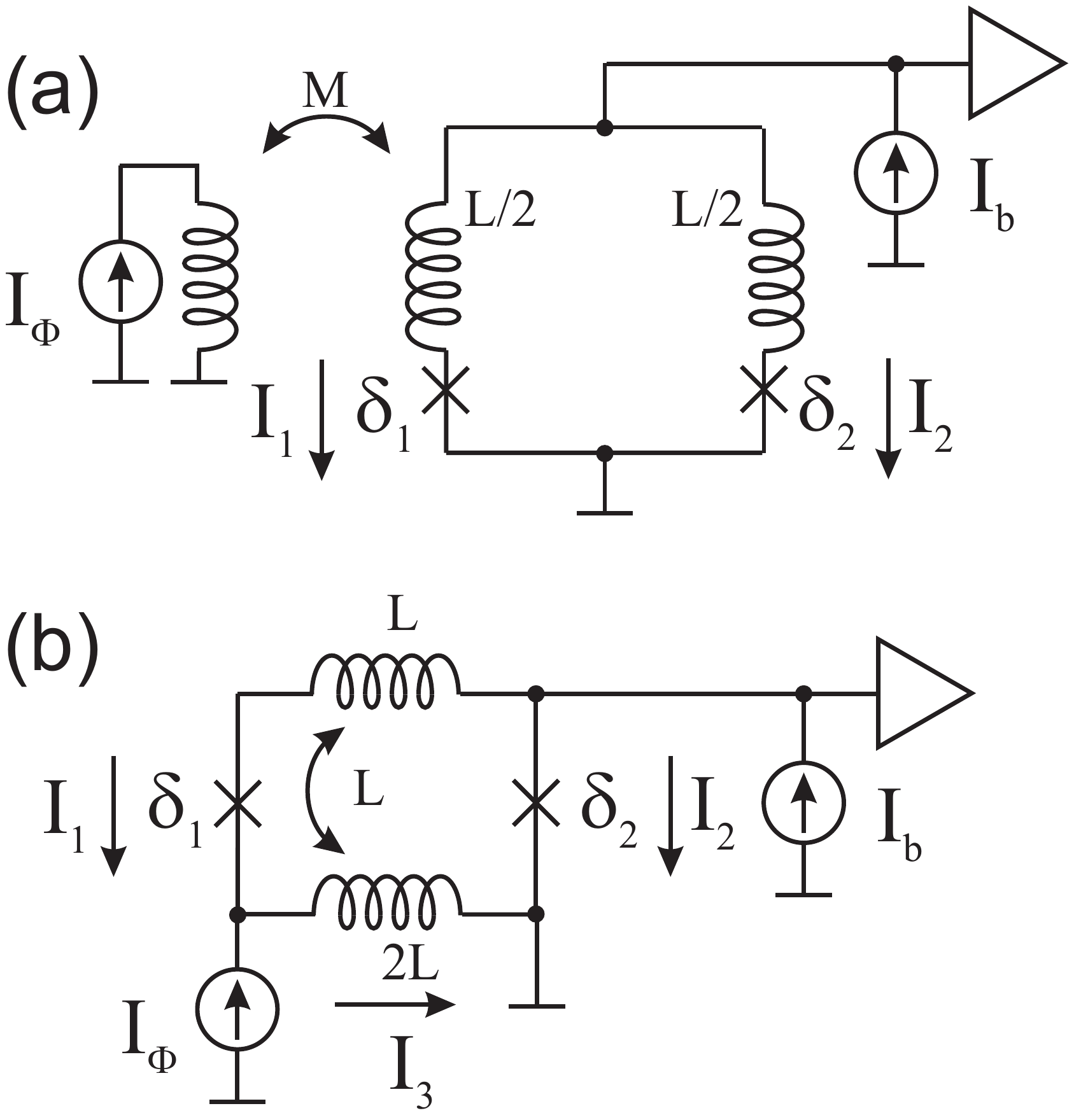}
\vspace*{-0.0in} \caption{Device geometries. (a) Symmetric dc SQUID. (b) Symmetric SLUG.}
\label{fig:circuits}
\end{center}
\end{figure}

To make contact with the earlier numerical studies of Tesche and Clarke \cite{Tesche77}, we begin by considering the familiar symmetric dc SQUID, shown in Fig. \ref{fig:circuits}a. The gain element consists of two overdamped Josephson junctions embedded in a superconducting loop with inductance $L$. The junctions (with gauge invariant phases $\delta_{1,2}$) have equal critical currents $I_0$, self-capacitances $C$, and shunt resistances $R$. The superconducting loop is formed from two equal branches with inductance $L/2$; we neglect the mutual inductance between the branches. A dc bias current $I_b$ and bias flux $\Phi_b$ establish a quasistatic operating point, and the signal is injected into an input coil that is coupled to the SQUID loop with mutual inductance $M$. The currents through the junctions are given by
\begin{align}
I_1 &= I_0\sin{\delta_1} + \frac{(V_1 - V_{n,1})}{R} + C \, \frac{dV_1}{dt} 
\nonumber \\
I_2 &= I_0\sin{\delta_2} + \frac{(V_2 - V_{n,2})}{R} + C \, \frac{dV_2}{dt},
\end{align}
where $V_{n,1}$ and $V_{n,2}$ are noise voltages associated with the resistive shunts, and where the voltages $V_{1,2}$ are related to the junction phases by the ac Josephson relation:
\begin{align}
V_1 &= \frac{\Phi_0}{2 \pi} \frac{d \delta_1}{dt} 
\nonumber \\
V_2 &= \frac{\Phi_0}{2 \pi} \frac{d \delta_2}{dt}.
\end{align}
Here, $\Phi_0~=~h/2e$ is the magnetic flux quantum. The SQUID loop supports a circulating current $J$ given by
\begin{align}
J = \frac{I_1 - I_2}{2}.
\end{align}
The voltage across the device is given by
\begin{align}
V &= V_1 + \frac{L}{2} \, \frac{dI_1}{dt} 
\nonumber \\
&= V_2 + \frac{L}{2} \, \frac{dI_2}{dt}.
\end{align}
The circulating current and the junction phases are related to the total flux in the loop $\Phi_T$ as follows:
\begin{align}
\Phi_T &= \Phi_b + LJ
\nonumber \\
&= \frac{\Phi_0}{2 \pi} (\delta_2 - \delta_1).
\end{align}
\begin{figure}[t]
\begin{center}
\includegraphics[width=.49\textwidth]{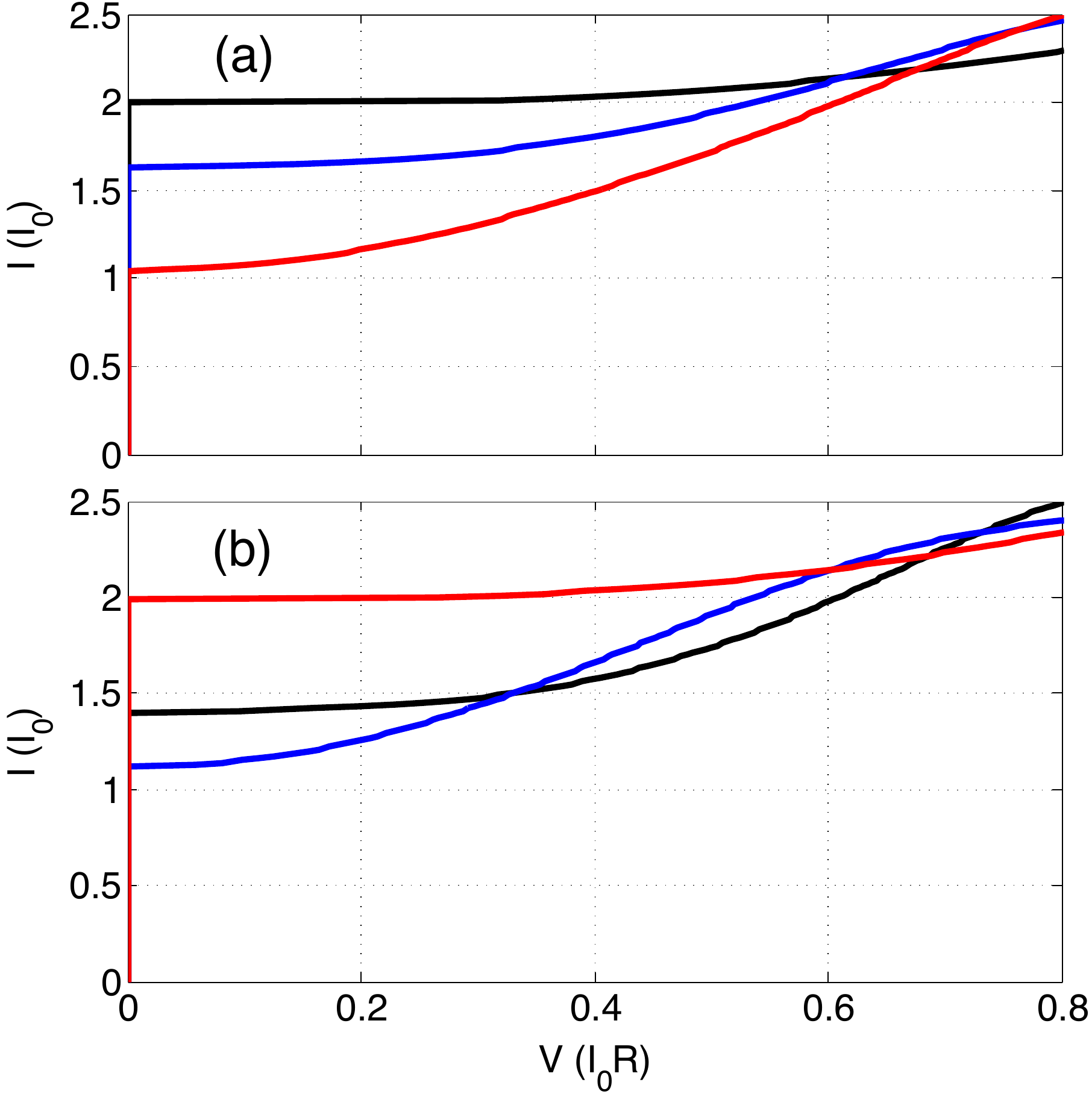}
\vspace*{-0.0in} \caption{I-V characteristics of (a) symmetric dc SQUID and (b) symmetric SLUG for $\Phi_b = 0 \, \Phi_0$ (black), $ 0.25 \, \Phi_0$ (blue), and $0.5 \, \Phi_0$ (red). The device parameters are $\beta_L = 1$ and $\beta_C = 0.8$.}
\label{fig:IV}
\end{center}
\end{figure}
\begin{figure}[t]
\begin{center}
\includegraphics[width=.49\textwidth]{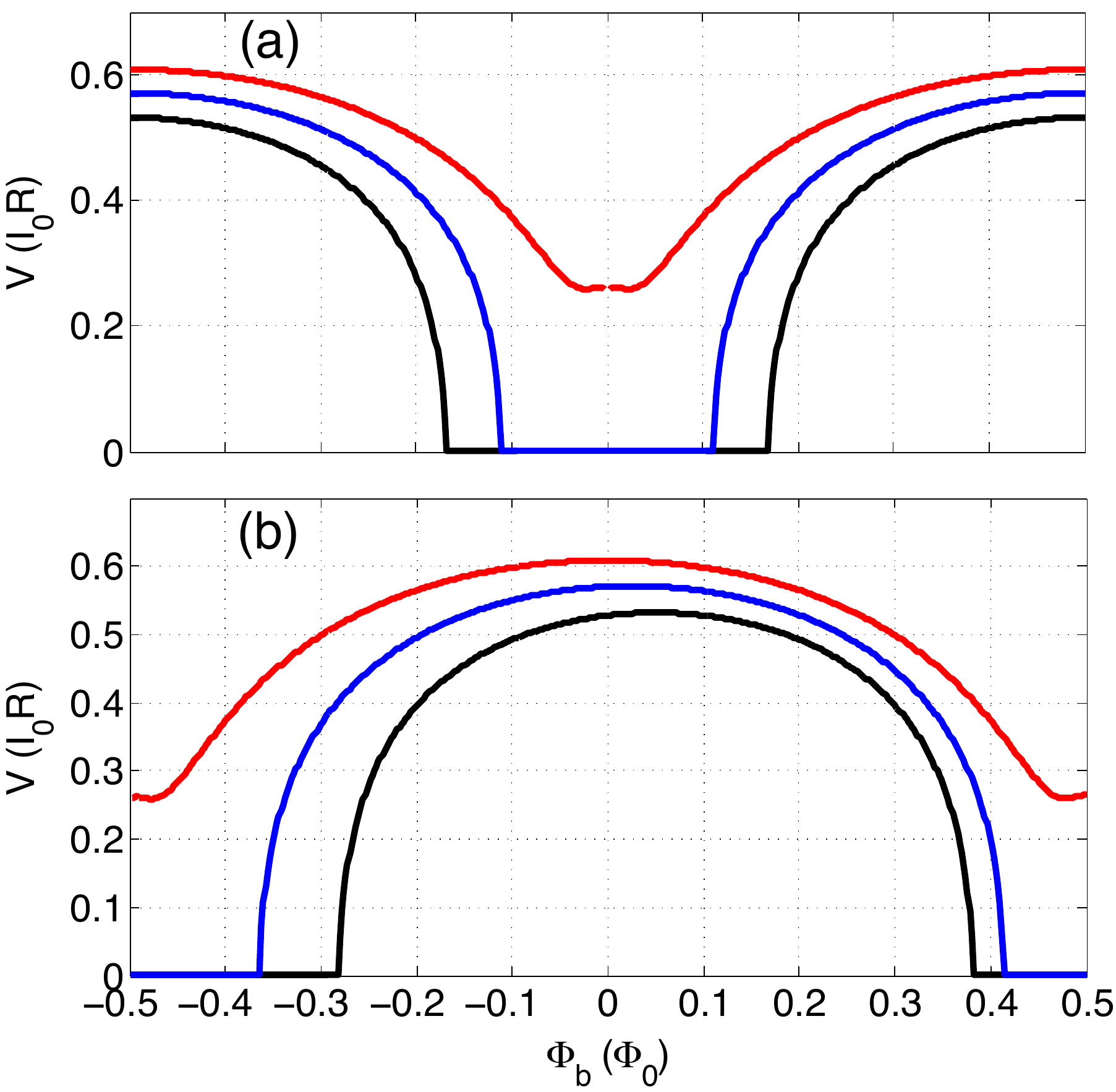}
\vspace*{-0.0in} \caption{V-$\Phi$ characteristics of (a) symmetric dc SQUID and (b) symmetric SLUG for $I_b = 1.8 \, I_0$ (black), $ 1.9 \, I_0$ (blue), and $2.0 \, I_0$ (red). The device parameters are $\beta_L = 1$ and $\beta_C = 0.8$.}
\label{fig:VPhi}
\end{center}
\end{figure}

We introduce dimensionless variables $i$, $v$, $\phi$, and $\theta$, defined as follows: $i~\equiv~I/I_0$, $v~\equiv~V/I_0 R$, $\phi~\equiv~\Phi/\Phi_0$, and $\theta~\equiv t~/~[\Phi_0/(2 \pi I_0 R)]$. In addition, we introduce the dimensionless reduced inductance $\beta_L = 2 I_0 L/\Phi_0$ and the damping parameter $\beta_C = (2 \pi/\Phi_0) I_0 R^2 C$. The equations of motion for the junction phases are written as
\begin{align}
\beta_C \ddot{\delta_1} &= \frac{i_b}{2} + \frac{\delta_2 - \delta_1 - 2\pi \phi_b}{\pi \beta_L} - \sin{\delta_1} - \dot{\delta_1} + v_{n,1} \nonumber \\
\beta_C \ddot{\delta_2} &= \frac{i_b}{2} - \frac{\delta_2 - \delta_1 - 2\pi \phi_b}{\pi \beta_L} - \sin{\delta_2} - \dot{\delta_2}  + v_{n,2}.
\label{eqn:SQUID}
\end{align}

The quasistatic output voltage and circulating current are given by
\begin{align}
v_{out} &= \frac{1}{2}\left( \dot{\delta_1} + \dot{\delta_2} \right) \\
j &= \frac{1}{\pi \beta_L}\left( \delta_2 - \delta_1 - 2\pi \phi_b \right).
\end{align}

In the SLUG geometry of Fig. \ref{fig:circuits}b, the device loop is formed from two superconducting traces separated by a thin dielectric layer, and the input signal is injected directly into one of the traces. In the case where the SLUG is integrated into a microstrip transmission line, the device is realized in three metallization steps (corresponding to the circuit groundplane and the two arms of the SLUG), with two dielectric thin films separating the metal layers. The mutual inductance between the arms of the device is of order the self-inductance of the arms, and must be taken into account. Below for the sake of simplicity we consider the case where the two dielectric films separating the superconducting layers are of equal thickness, resulting in branch inductances $2L$ and $L$ with mutual inductance $L$. The total inductance of the device is then $L$, and the mutual coupling $M$ of the input current $I_\Phi$ to the device loop is also $L$. We refer to this configuration as the \textit{symmetric SLUG}. 
The total flux through the device becomes
\begin{align}
\Phi_T &= L\left(I_1 + I_\Phi \right) + \Phi_b
\nonumber \\
	   &= \frac{\Phi_0}{2\pi}\left(\delta_2 - \delta_1\right).
\end{align}	
We write the dimensionless equations of motion for $\delta_{1,2}$ as follows:
\begin{align}
\beta_C \ddot{\delta_1} &= \frac{\delta_2 - \delta_1 - 2\pi \phi_b}{\pi \beta_L} - i_\phi - \sin{\delta_1} - \dot{\delta_1} + v_{N,1} \nonumber \\
\beta_C\ddot{\delta_2} &= - \frac{\delta_2 - \delta_1 - 2\pi \phi_b}{\pi \beta_L} + i_b + i_\phi - \sin{\delta_2} - \dot{\delta_2}  + v_{N,2}.
\label{eqn:SLUG}
\end{align}
%
The output voltage and circulating current are given by
\begin{align}
v_{out} &= \dot{\delta_2} \\
j &= \frac{1}{\pi \beta_L}\left( \delta_2 - \delta_1 - 2\pi \phi_b \right) - i_\phi/2.
\end{align}
To operate the SQUID or the SLUG as an amplifier, one chooses $I_b$ and $\Phi_b$ to establish a quasistatic operating point where the transfer function $V_\Phi \equiv \partial V/\partial \Phi$ is large. In both cases, the device acts as a transimpedance element: the input signal is coupled to the device as a current, and the output signal is coupled from the device as a voltage.

\section{III. dc Characteristics}

Equations \eqref{eqn:SQUID} and \eqref{eqn:SLUG} were numerically integrated using a 4th order Runge-Kutta solver for $N \sim 2^{18}$ time steps $\Delta \theta$ over a range of bias points. 
In Figs. \ref{fig:IV}a-b we show the I-V characteristics of the symmetric dc SQUID and the symmetric SLUG with $\beta_L = 1$ and $\beta_C = 0.8$; in Figs. \ref{fig:VPhi}a-b we show the V-$\Phi$ characteristics of the same devices. For bias near $1.9 \, I_0$, the peak-to-peak voltage modulation is around 0.5~$I_0 R$.
\begin{figure}[t]
\begin{center}
\includegraphics[width=.49\textwidth]{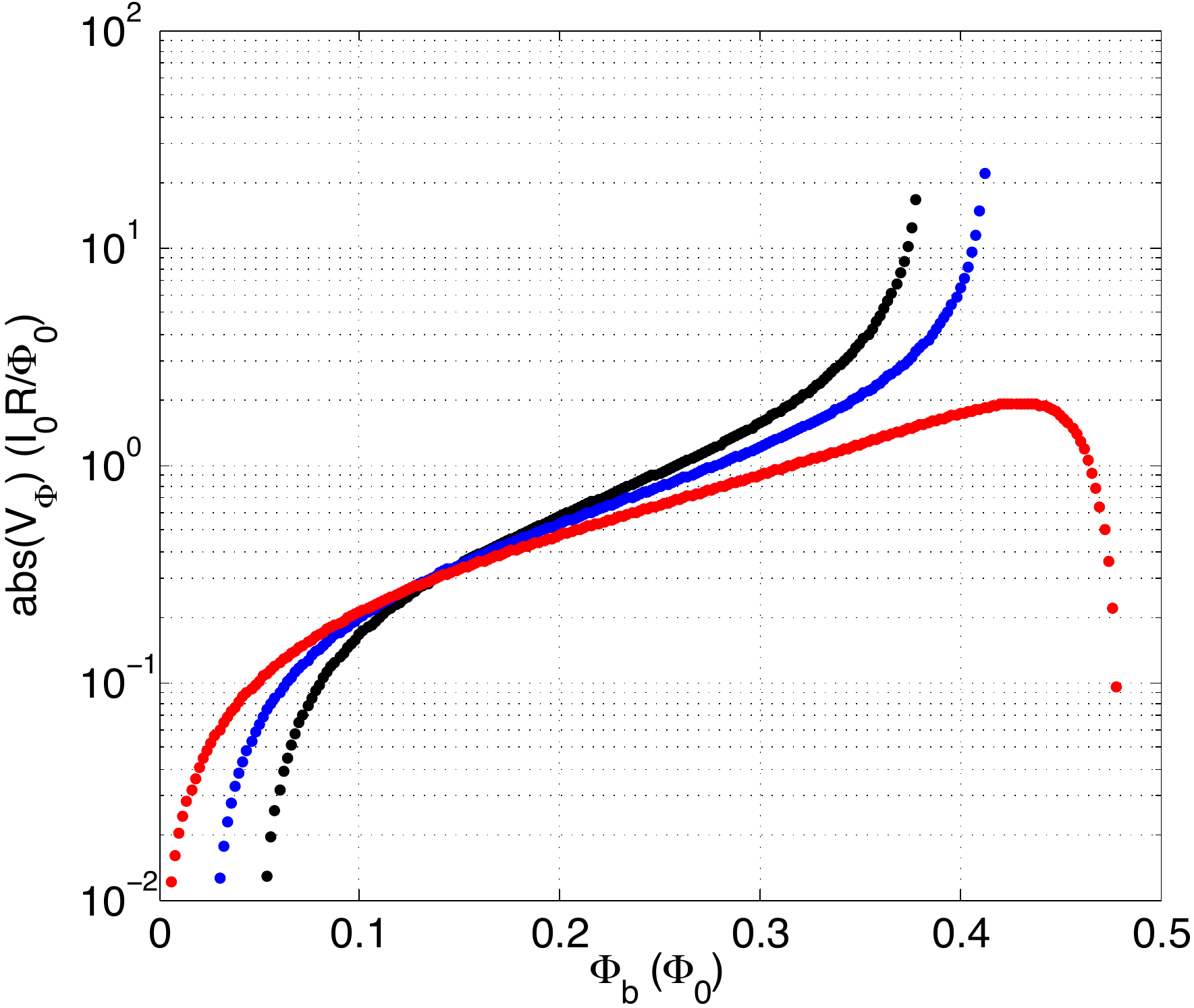}
\vspace*{-0.0in} \caption{Forward transfer function $V_\Phi$ of SLUG circuit \textit{versus} quasistatic bias flux for bias currents $I_b = 1.8 \, I_0$ (black), $ 1.9 \, I_0$ (blue), and $2.0 \, I_0$ (red). The device parameters are $\beta_L = 1$ and $\beta_C = 0.8$.}
\label{fig:dVdPhi}
\end{center}
\end{figure}
We observe that the dc characteristics of the SLUG closely match those of the symmetric dc SQUID, apart from a shift in flux bias point that arises from the asymmetric division of the SLUG bias current. Similarly, we have found that the scattering parameters and noise properties of the SLUG and the SQUID are closely matched, apart from this bias shift. Therefore for the sake of simplicity we choose to focus in the following on the device characteristics of the SLUG alone.
We will consider the following set of SLUG parameters: $\beta_L=1$, $\beta_C = 0.8$, $L$~=~10~pH, and $C$~=~50~fF, corresponding to a junction with critical current 100 $\mu$A and area around 1~$\mu$m$^2$. Several considerations lead us to this choice. First, inductances of order 10~pH are realized in a reliable, controlled way using the SLUG geometry, and the resulting device is immune from stray reactances and straightforward to model at microwave frequencies. The required critical current density is 10~kA/cm$^2$, within the reach of standard Nb-AlO$_x$-Nb technology. While Joule heating in the shunt resistors is significant, the addition of large-volume normal-metal cooling fins should allow equilibration of the shunt resistors at temperatures below 100~mK (see Section X). Lower device inductances would require uncomfortably high junction critical currents to achieve comparable device performance, and fabrication yield and Joule heating of the shunts would become problematic. On the other hand, a significantly larger SLUG inductance would provide less gain and complicate the microwave engineering, owing to the larger device dimensions.

\section{IV. Scattering Parameters}

In order to optimize SLUG amplifier design, it is necessary to understand the forward transfer function and the complex input and output impedances of the device. To extract these from our model, we apply appropriate test signals and probe the complex response at the excitation frequency, chosen to be a small fraction of the Josephson frequency $\omega_J/2 \pi$. The forward transimpedance $V_I \equiv \partial V/\partial I$ is readily derived from the SLUG flux-to-voltage transfer function $V_\Phi$:
\begin{align}
V_I = M V_\Phi,
\end{align}
where again we have $M=L$ for the case of the symmetric SLUG. In Fig. \ref{fig:dVdPhi} we plot SLUG $V_\Phi$ \textit{versus} flux over a range of current bias points for $\beta_L = 1$ and $\beta_C = 0.8$.
\begin{figure}[t]
\begin{center}
\includegraphics[width=.49\textwidth]{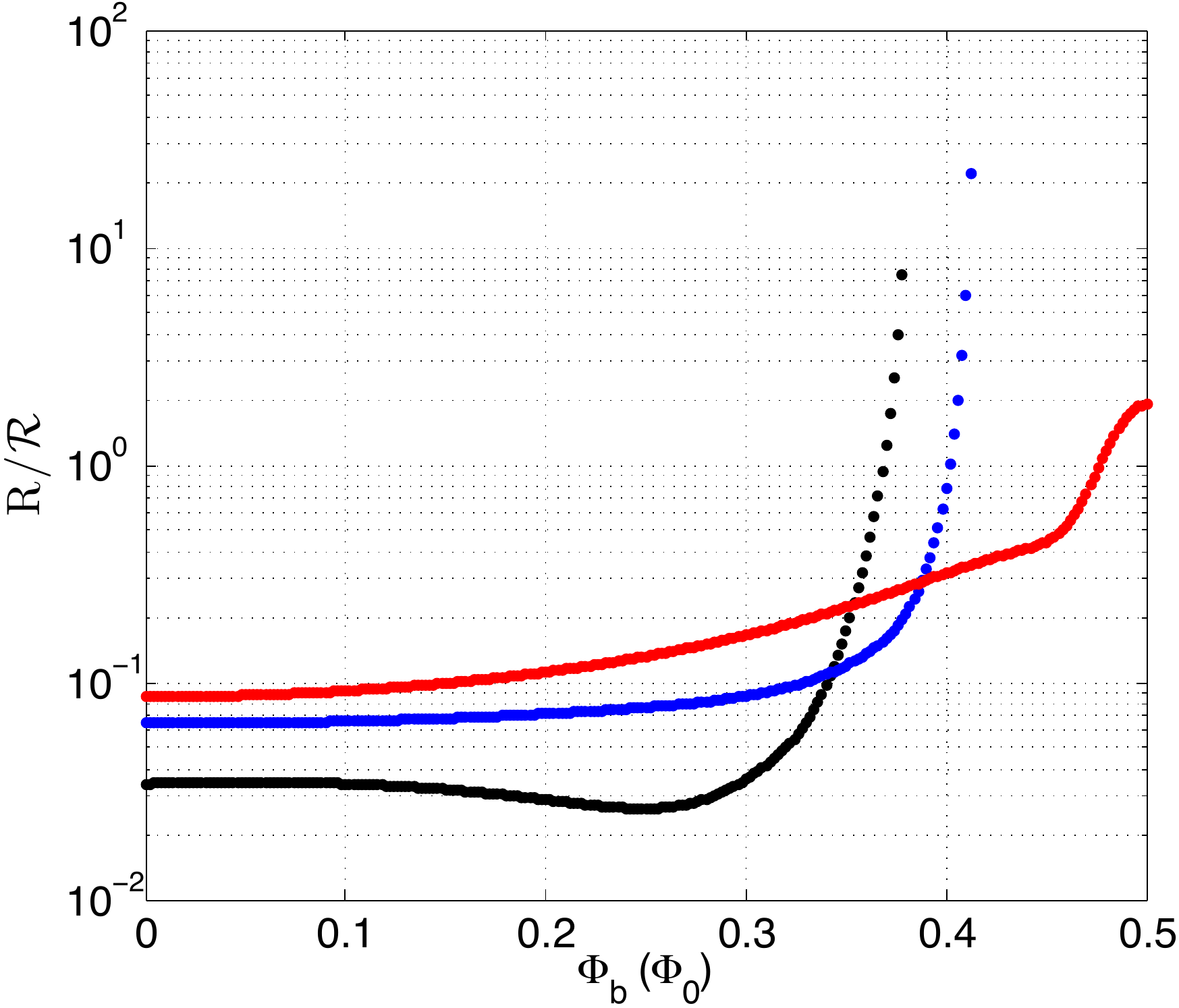}
\vspace*{-0.0in} \caption{$R/\mathcal{R}$ \textit{versus} flux for a SLUG with $\beta_L = 1$ and $\beta_C = 0.8$, for bias currents $I_b = 1.8 \, I_0$ (black), $ 1.9 \, I_0$ (blue), and $2.0 \, I_0$ (red).}
\label{fig:Rin}
\end{center}
\end{figure}

\begin{figure}[t]
\begin{center}
\includegraphics[width=.49\textwidth]{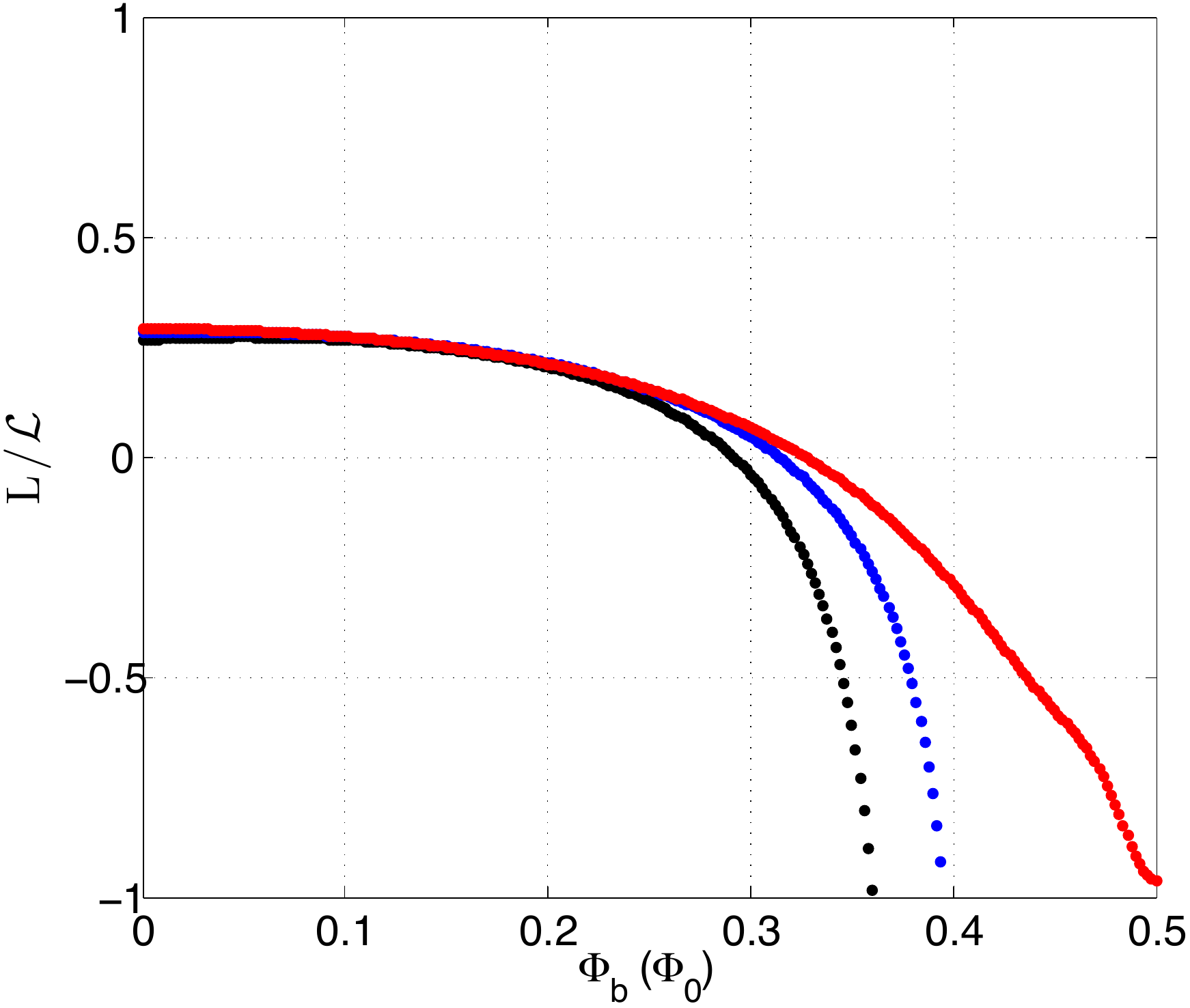}
\vspace*{-0.0in} \caption{$L/\mathcal{L}$ \textit{versus} flux for a SLUG with $\beta_L = 1$ and $\beta_C = 0.8$, for bias currents $I_b = 1.8 \, I_0$ (black), $ 1.9 \, I_0$ (blue), and $2.0 \, I_0$ (red).}
\label{fig:Xin}
\end{center}
\end{figure}

Next we consider the input return loss. The SLUG input is an inductive short to ground at low frequencies, and the complex input impedance $Z_i$ is frequency dependent. The input impedance is readily derived from the dynamic impedance $\mathcal{Z}$, defined in terms of the flux-to-current transfer function $J_\Phi \equiv \partial J/\partial \Phi$ as follows:
\begin{align}
-J_\Phi \equiv \frac{1}{\mathcal{Z}} = \frac{1}{\mathcal{L}} + \frac{j \omega}{\mathcal{R}},
\label{eqn:Zdyn}
\end{align}
where following Hilbert \textit{et al.} we have introduced the frequency-independent dynamic resistance $\mathcal{R}$ and dynamic inductance $\mathcal{L}$ \cite{Hilbert85a}. 
In Figs. \ref{fig:Rin} and \ref{fig:Xin} we plot $R/\mathcal{R}$ and $L/\mathcal{L}$, respectively, for a SLUG with $\beta_L = 1$ and $\beta_C = 0.8$ over a range of bias points.
\begin{figure}[t]
\begin{center}
\includegraphics[width=.49\textwidth]{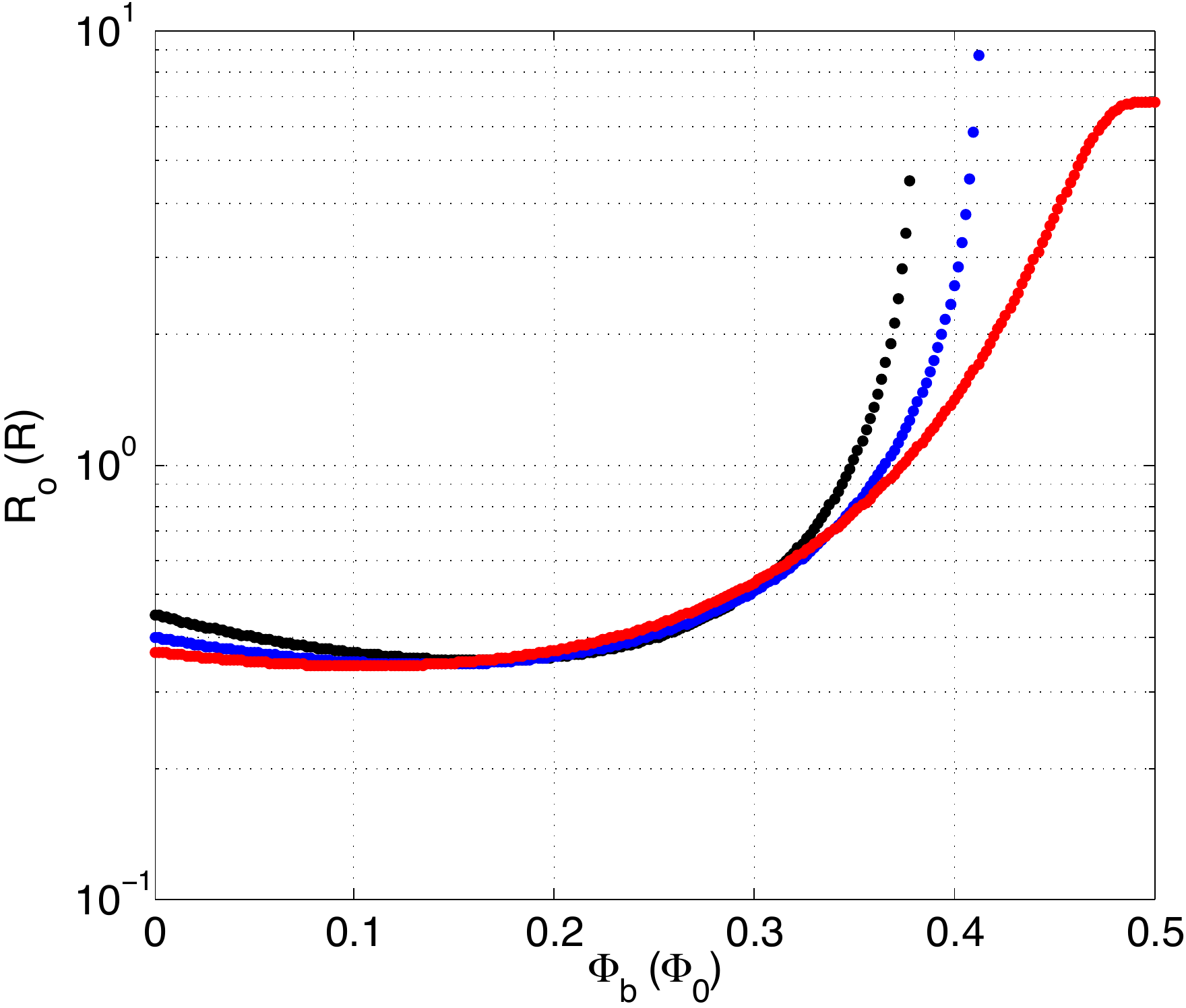}
\vspace*{-0.0in} \caption{SLUG output resistance $R_o$ \textit{versus} flux for bias currents $I_b = 1.8 \, I_0$ (black), $ 1.9 \, I_0$ (blue), and $2.0 \, I_0$ (red). The device parameters are $\beta_L = 1$ and $\beta_C = 0.8$.}
\label{fig:Rout}
\end{center}
\end{figure}

Finally, in Fig. \ref{fig:Rout} we show the device output impedance $R_o$ over a range of bias points. The output impedance is real and frequency independent, and the magnitude of $R_o$ is of order the junction shunt resistance $R$. 
\begin{figure}[t]
\begin{center}
\includegraphics[width=.47\textwidth]{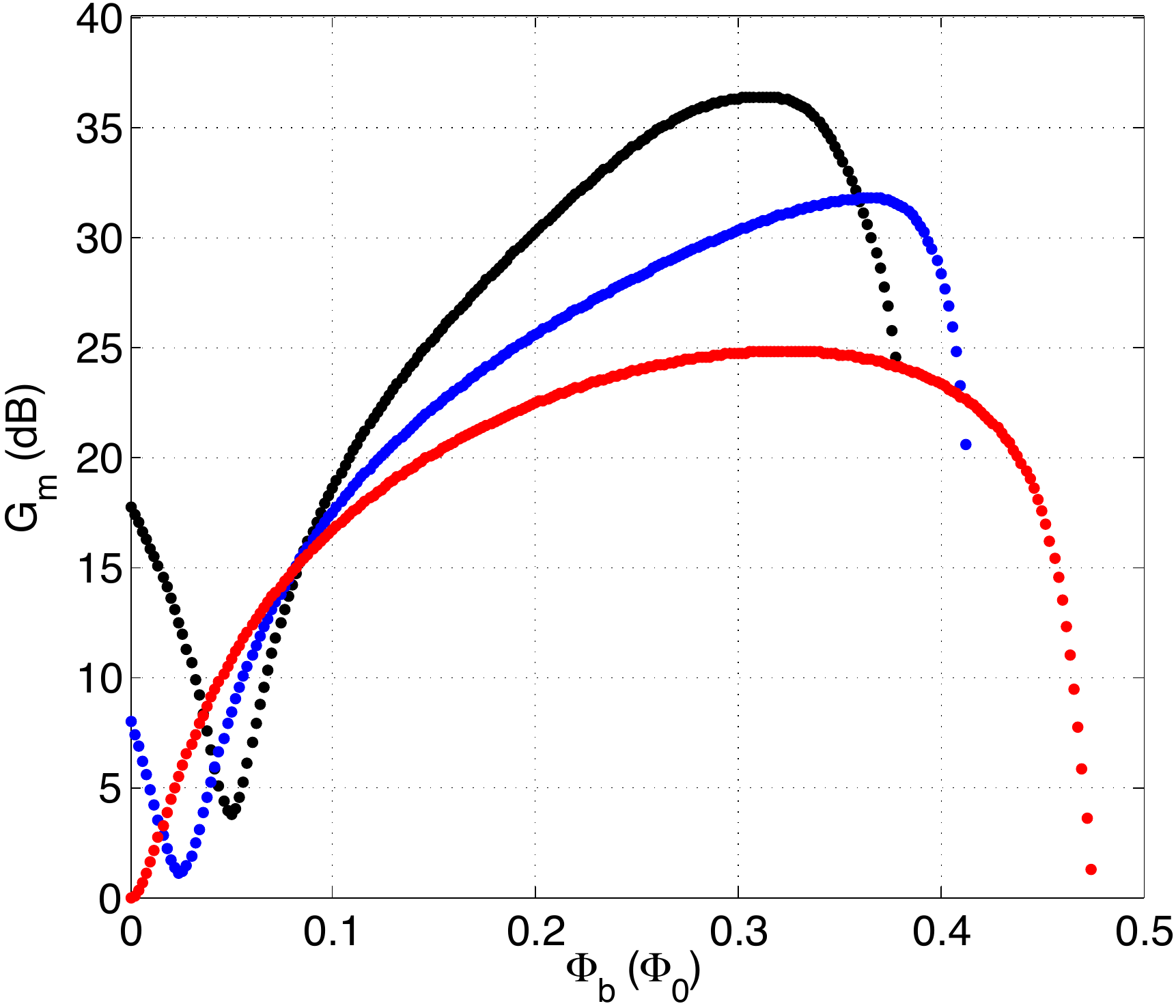}
\vspace*{-0.0in} \caption{Maximum achievable power gain $G_m$ for a SLUG amplifier \textit{versus} flux for bias currents $I_b = 1.8 \, I_0$ (black), $ 1.9 \, I_0$ (blue), and $2.0 \, I_0$ (red). The device parameters are $\beta_L = 1$, $\beta_C = 0.8$, $L$~=~10~pH, and $C$~=~50~fF; the operating frequency is 5~GHz.}
\label{fig:gain}
\end{center}
\end{figure}

For the following discussion, it is convenient to work in terms of the bias-dependent dimensionless impedance parameters $\rho_{i,o}$, defined as follows:
\begin{align}
R_i &= \rho_i \frac{(\omega M)^2}{R}
\nonumber \\
R_o &= \rho_o R
\label{eqn:rho}
\end{align}
From the definition of $\mathcal{R}$ it follows that $\rho_i = R/\mathcal{R}$. As we will see, amplifier gain, bandwidth, and noise properties depend sensitively on $\rho_i$ and $\rho_o$.

Power gain of the device is maximized when appropriate conjugate matching networks are employed to couple the signal to and from the device. The maximum available power gain $G_m$ is given as follows:
\begin{align}
G_m &= \frac{V_o^2/4 R_o}{I_i^2 R_i} \\
\nonumber
\end{align}
where $I_i$ is the input current and $V_o$ is the output voltage. Using Eq. \eqref{eqn:rho}, we find
\begin{align}
G_m &= \frac{1}{4 \rho_i \rho_o} \left(\frac{V_\Phi}{\omega}\right)^2.
\end{align}

In Fig. \ref{fig:gain} we plot $G_m$ for the symmetric SLUG with $\beta_L = 1$, $\beta_C = 0.8$, $L$~=~10~pH, and $C$~=~50~fF for an operating frequency of 5~GHz. Over a broad range of bias parameters gain in excess of 20 dB is readily achievable. It is important to note, however, that a conjugate match to a 50~$\Omega$ source does not yield best amplifier noise performance, due to the mismatch between the real part of the SLUG input impedance $R_i$ and the optimal noise-matched source impedance, which can be significantly larger than $R_i$. Amplifier optimization therefore involves a tradeoff between gain and noise performance, as discussed in detail below.
\begin{figure}[t]
\begin{center}
\includegraphics[width=.47\textwidth]{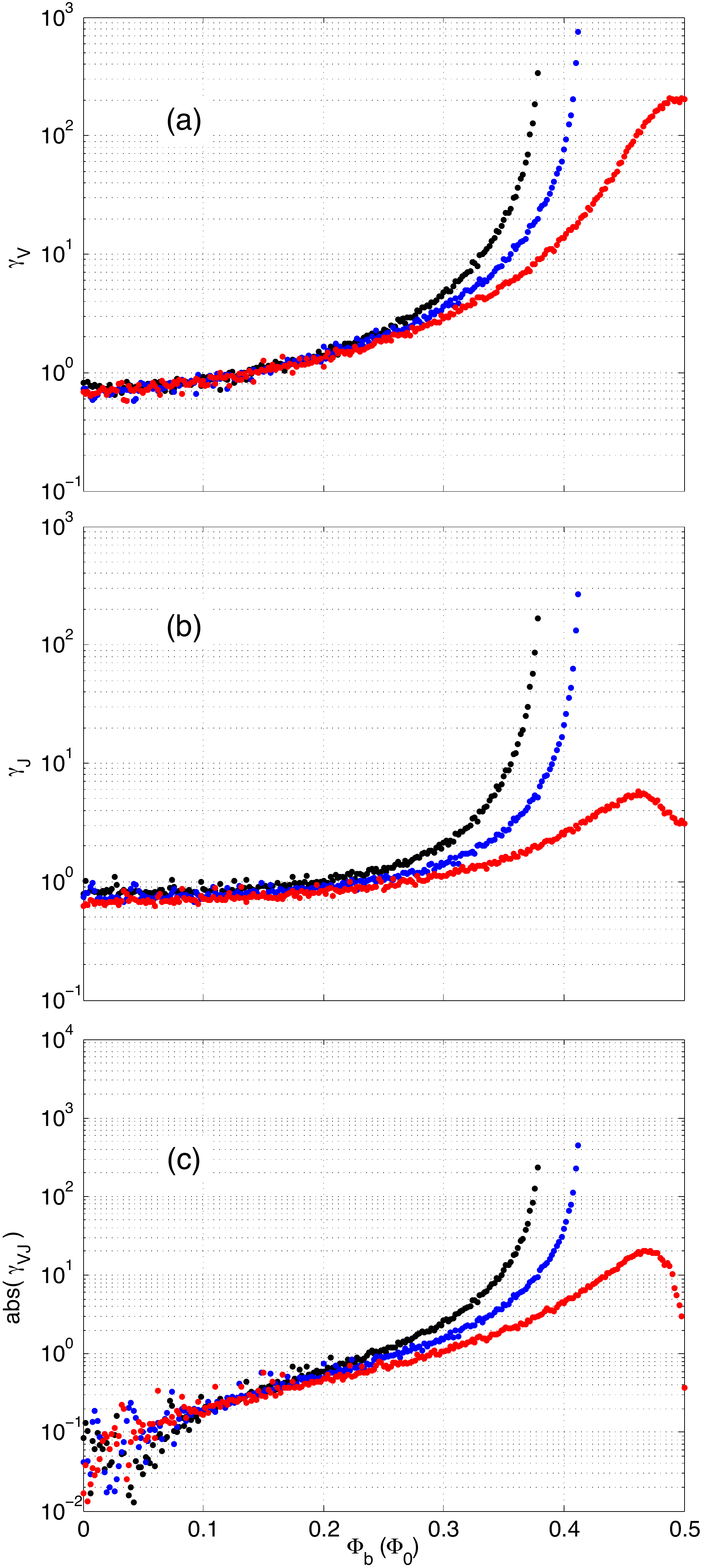}
\vspace*{-0.0in} \caption{Dimensionless SLUG noises (a) $\gamma_V$, (b) $\gamma_J$, and (c) $\gamma_{VJ}$ \textit{versus} flux for bias currents $I_b = 1.8 \, I_0$ (black), $ 1.9 \, I_0$ (blue), and $2.0 \, I_0$ (red). The SLUG parameters are $\beta_L = 1$, $\beta_C = 0.8$, and $\Gamma = 4 \times 10^{-5}$.}
\label{fig:gammas}
\end{center}
\end{figure}

The bandwidth of the SLUG amplifier will be determined by the coupling to the low-impedance input port, as the device output is reasonably well-matched to typical transmission line impedances. To get a rough idea of amplifier bandwidth we consider a 50~$\Omega$ source impedance and assume that conjugate matching at the device input is accomplished via a simple quarter-wave transmission line section; for simplicity we neglect the imaginary part of the SLUG input impedance. The amplifier quality factor $Q$ is given by
\begin{align}
Q &\approx \frac{\pi}{8}\sqrt{\frac{50 \, \Omega}{R_i}}
\nonumber \\
  &= \frac{\pi}{8 \omega M} \sqrt{\frac{50 \, \Omega \times R}{\rho_i}}.
  \label{eqn:Q}
\end{align}
\begin{figure}[t]
\begin{center}
\includegraphics[width=.49\textwidth]{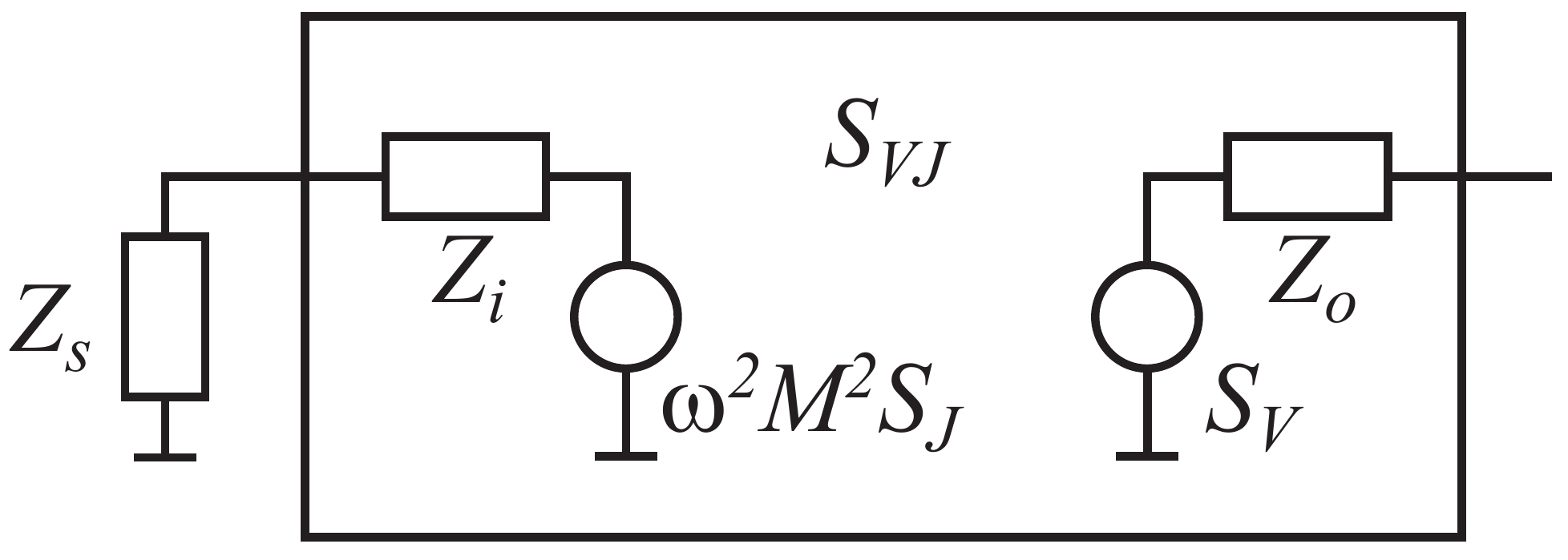}
\vspace*{-0.0in} \caption{Circuit for noise analysis.}
\label{fig:noise_analysis}
\end{center}
\end{figure}
\begin{figure}[t]
\begin{center}
\includegraphics[width=.49\textwidth]{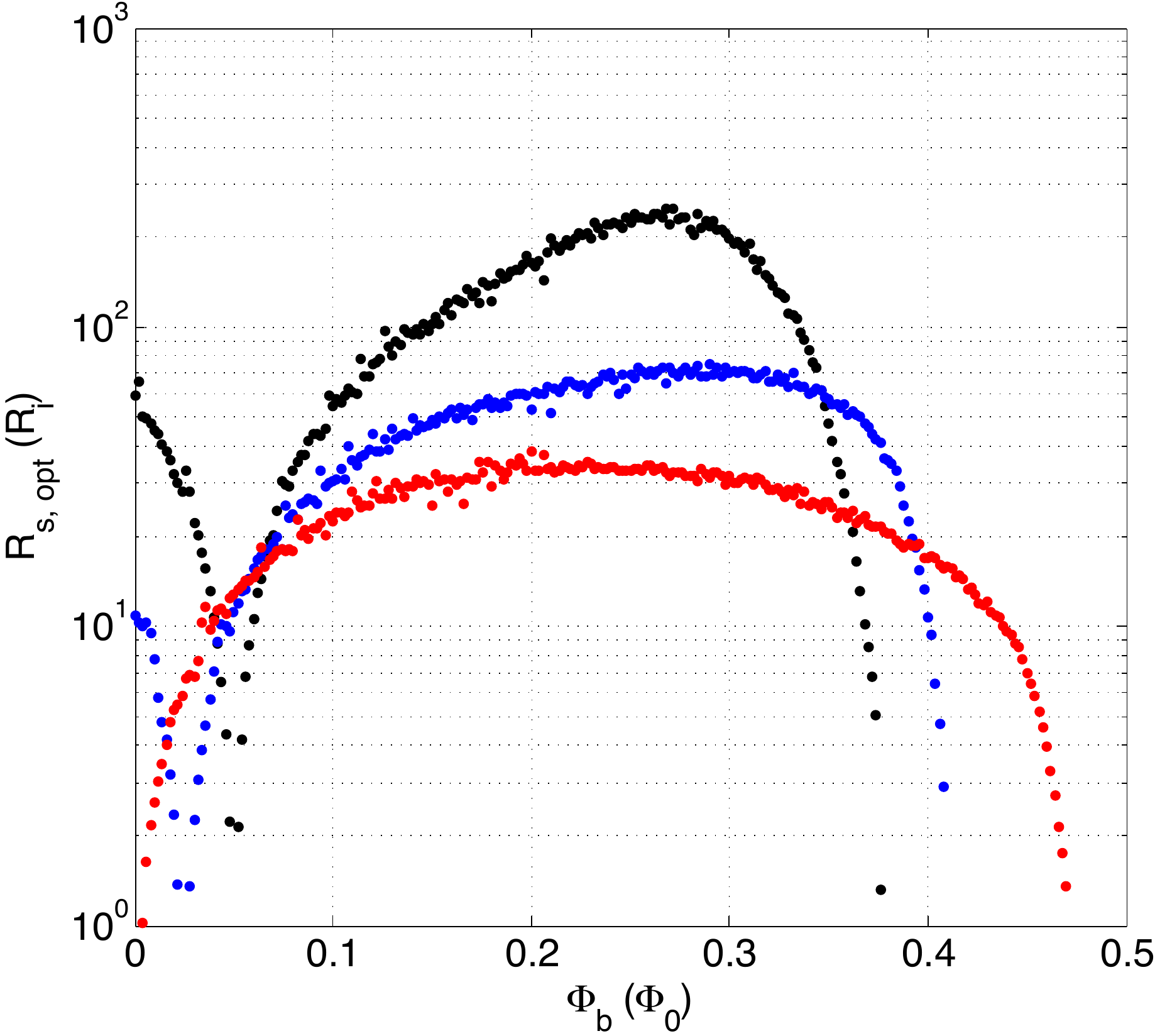}
\vspace*{-0.0in} \caption{Real part of the optimal source impedance $R_{s,opt}$ \textit{versus} flux for bias currents $I_b = 1.8 \, I_0$ (black), $ 1.9 \, I_0$ (blue), and $2.0 \, I_0$ (red). The SLUG parameters are $\beta_L = 1$, $\beta_C = 0.8$, and $\Gamma = 4 \times 10^{-5}$. The operating frequency is 5~GHz.}
\label{fig:Rs_Ri}
\end{center}
\end{figure}
The bandwidth of an amplifier designed at an operating frequency $\omega/2 \pi$ is then $\omega/2 \pi Q$. For an operating frequency around 5~GHz, we find that $R_i$ is of order 0.1~$\Omega$. Therefore we expect $Q$ of order 10, and amplifier bandwidths of order hundreds of MHz. For current bias $I_b < 2 I_0$ and for a narrow range of fluxes corresponding to bias points near the supercurrent branch, we find that it is possible to achieve extremely high power gain (see Fig. \ref{fig:gain}). However, the high gains achieved at these bias points are due largely to vanishing $R_i$; an amplifier designed to operate in this regime would have a rather small bandwidth. It is important to note that Eq. \eqref{eqn:Q} presents only a rough guideline for the bandwidth rather than a fundamental limit. In particular, it is possible to obtain a larger bandwidth with no degradation in gain by employing either a tapered transmission line matching section or a multisection input transformer with stepped transmission line impedances. We postpone a more detailed discussion of amplifier bandwidth to Section VII.

\section{V. Noise Properties in the Thermal Regime}

The Johnson noise of the shunt resistors gives rise to a voltage noise at the device output and to a circulating current noise in the device loop; moreover, these noises are partially correlated, since the circulating current noise couples a flux noise to the loop, which in turn yields a voltage noise across the device. To incorporate noise in our model, we used a pseudorandom number generator to create a gaussian-distributed set of voltages $v_{N,1}$ and $v_{N,2}$ with zero mean and variance $2\Gamma/\Delta \theta$, where we have introduced the dimensionless noise parameter $\Gamma = 2\pi k_B T/I_0 \Phi_0$; this choice corresponds to the usual white power spectral density $S_v = 4\Gamma$ for Johnson noise in the thermal limit. The simulations were averaged over many ($\sim$~100) realizations of the random noise voltages. Following Clarke \textit{et al.}, we introduce the dimensionless noise parameters $\gamma_V$, $\gamma_J$, and $\gamma_{VJ}$, such that the voltage noise spectral density at the device output is given by $S_V=2 \gamma_V k_B T R$, the circulating current noise spectral density is $S_J = 2 \gamma_J k_B T/R$, and the cross noise spectral density is $S_{VJ} = 2 \gamma_{VJ} k_B T$; here $T$ is the electron temperature of the shunt resistors \cite{Tesche79, Hilbert85b}. These noises are calculated by solving the Langevin equations \eqref{eqn:SLUG}. The noise spectrum consists of a series of peaks at the Josephson frequency and its harmonics; the dimensionless noises $\gamma$ are evaluated at low frequency $f \ll \omega_J/2 \pi$ where the spectrum is white. The noises $\gamma$ do depend on the noise parameter $\Gamma$, due to the possibility of saturation and smearing of the device characteristics at elevated temperature. In Fig. \ref{fig:gammas} we plot the dimensionless noises over a range of bias parameters of the symmetric SLUG for $\beta_L = 1$, $\beta_C = 0.8$, and $\Gamma = 4 \times 10^{-5}$; this choice corresponds to a temperature of 100 mK for a junction critical current of 100 $\mu$A. We note that at high bias current $I_b \gg I_0$, $\gamma_{V, J}$ approach the expected Johnson noise limit of $1$ for the two shunt resistors in parallel.

The device noise temperature $T_n$ can be evaluated from the circuit shown in Fig. \ref{fig:noise_analysis}. We assume a noiseless source impedance $Z_s = R_s + j X_s$ and equate the total noise of the amplifier to the noise contribution from a source resistance $R_s$ at an effective temperature $T_n$. We refer all noises to the device output. We find
\begin{align}
&4 k_B T_n R_s \, \frac{V_\Phi^2 M^2}{R_t^2 + X_t^2} \, =
\nonumber \\
&2\gamma_V k_B T R \, + \, \frac{2 \gamma_J k_B T}{R} \, \frac{\omega^2 V_\Phi^2 M^4}{R_t^2 + X_t^2} \, + \, 4 \gamma_{VJ} k_B T \, \frac{\omega V_\Phi M^2 X_t}{R_t^2 + X_t^2}.
\end{align}
Here $R_t = R_s + R_i$ ($X_t = X_s + X_i$) is the sum of the real (imaginary) parts of the source impedance and the device input impedance. The noise temperature is thus given by
\begin{align}
T_n = \left[\frac{\gamma_V}{2}\, \frac{(R_t^2+X_t^2)R}{V_\Phi^2 M^2 R_s} \,+\, \frac{\gamma_J}{2} \,\frac{\omega^2 M^2}{R R_s} \,+\, \gamma_{VJ}\,\frac{\omega X_t}{V_\Phi R_s} \right] T .
\label{eqn:Tnthermal}
\end{align}
\begin{figure}[t]
\begin{center}
\includegraphics[width=.49\textwidth]{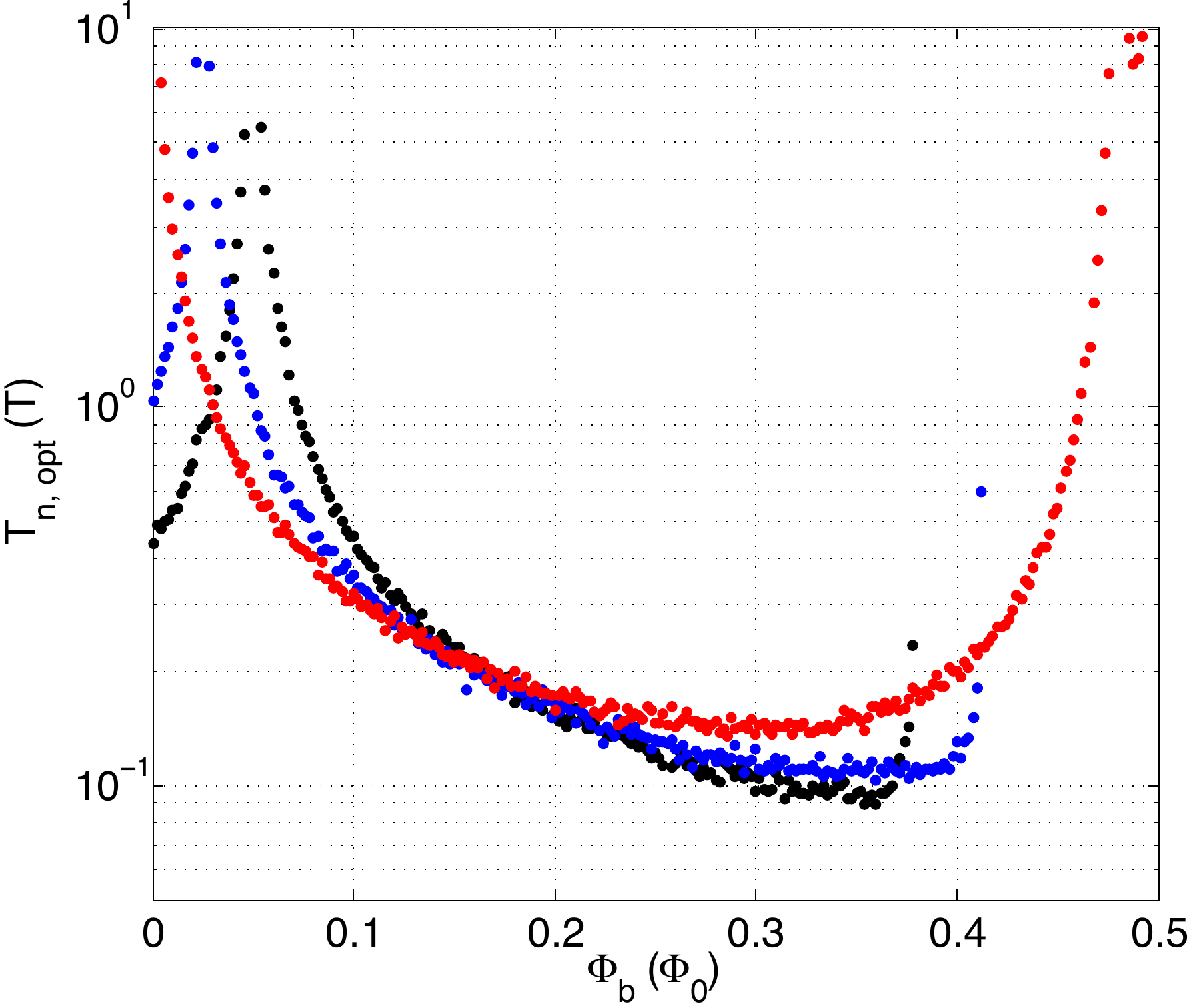}
\vspace*{-0.0in} \caption{Optimal SLUG noise temperature \textit{versus} flux for bias currents $I_b = 1.8 \, I_0$ (black), $ 1.9 \, I_0$ (blue), and $2.0 \, I_0$ (red). The SLUG parameters are $\beta_L = 1$, $\beta_C = 0.8$, and $\Gamma = 4 \times 10^{-5}$. The operating frequency is 5~GHz.}
\label{fig:Tn_thermal}
\end{center}
\end{figure}
We use the condition $\partial T_n / \partial X_t = 0$ to solve for the imaginary part of the optimal source impedance. We find
\begin{align}
X_{s,opt} = -\frac{\gamma_{VJ}}{\gamma_V} \, \frac{\omega V_\Phi M^2}{R} - X_i.
\end{align}
Similarly, the condition $\partial T_n / \partial R_s = 0$ yields the real part of the optimal source impedance. We have
\begin{align}
R_{s,opt} = \left[1 + \frac{1}{\gamma_V^2 \rho_i^2}\left(\frac{V_\Phi}{\omega}\right)^2 \left(\gamma_V \gamma_J - \gamma_{VJ}^2\right) \right]^{1/2} R_i.
\end{align}
For bias points where $V_\Phi$ is highest, we have the following approximate expression for $R_{s,opt}$:
\begin{align}
R_{s,opt} \, \approx& \, \frac{1}{\gamma_V \rho_i} \frac{V_\Phi}{\omega} \left(\gamma_V \gamma_J - \gamma_{VJ}^2 \right)^{1/2} R_i 
\nonumber \\
=& \, \frac{\omega V_\Phi M^2}{\gamma_V R} \left(\gamma_V \gamma_J - \gamma_{VJ}^2 \right)^{1/2}.
\end{align}
In Fig. \ref{fig:Rs_Ri} we plot $R_{s,opt}/R_i$ \textit{versus} flux for various bias currents. For typical device parameters, we have $R_{s,opt} \gg R_i$. For this reason, it is not possible to achieve a simultaneous power match and noise match. It is worthwhile to note, however, that the ratio $R_{s,opt}/R_i$ scales with frequency as $\omega^{-1}$, facilitating simultaneous attainment of high gain and good noise performance at higher operating frequencies.

When the signal is coupled to the device via a source with optimal impedance $R_{s,opt} + j X_{s,opt}$, the amplifier noise temperature becomes
\begin{align}
T_{n,opt} \, = \, \frac{\omega}{V_\Phi} \left(\gamma_V \gamma_J - \gamma_{VJ}^2 \right)^{1/2} T.
\end{align}
In Fig. \ref{fig:Tn_thermal} we show the optimal noise temperature $T_{n,opt}$ for a SLUG amplifier over a range of bias points at an operating frequency $\omega / 2 \pi = 5$~GHz. Note that every point in these plots corresponds to a different realization of the input matching network; in Section VII we will examine the bias- and frequency-dependent noise temperature of SLUG amplifiers operated with a fixed input network.

\section{VI. Noise Properties in the Quantum Regime}

\begin{figure}[t]
\begin{center}
\includegraphics[width=.49\textwidth]{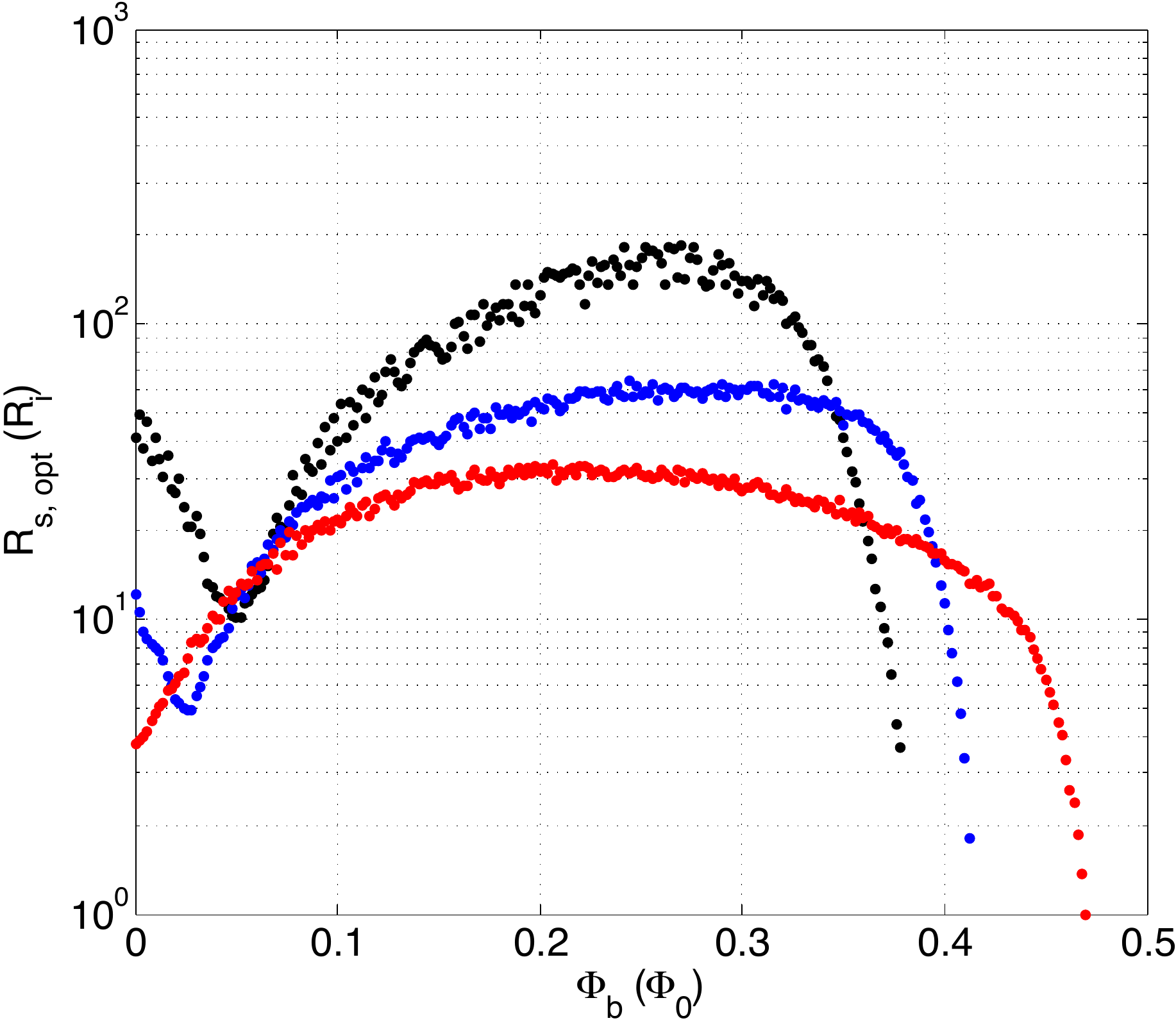}
\vspace*{-0.0in} \caption{Real part of the optimal source impedance $R_{s,opt}$ in the quantum regime \textit{versus} flux for bias currents $I_b = 1.8 \, I_0$ (black), $ 1.9 \, I_0$ (blue), and $2.0 \, I_0$ (red). The operating frequency is 5~GHz and the SLUG parameters are $\beta_L = 1$, $\beta_C = 0.8$, $L$~=~10~pH, and $C$~=~50~fF.}
\label{fig:Rsopt_quantum}
\end{center}
\end{figure}

At sufficiently low temperature, the zero-point fluctuations of the resistive shunts are expected to make the dominant noise contribution. The full expression for the spectral density of voltage noise produced by the resistors is written as $2 h f R \, \textrm{coth}(hf/2k_B T)$.
\begin{figure}[th!]
\begin{center}
\includegraphics[width=.47\textwidth]{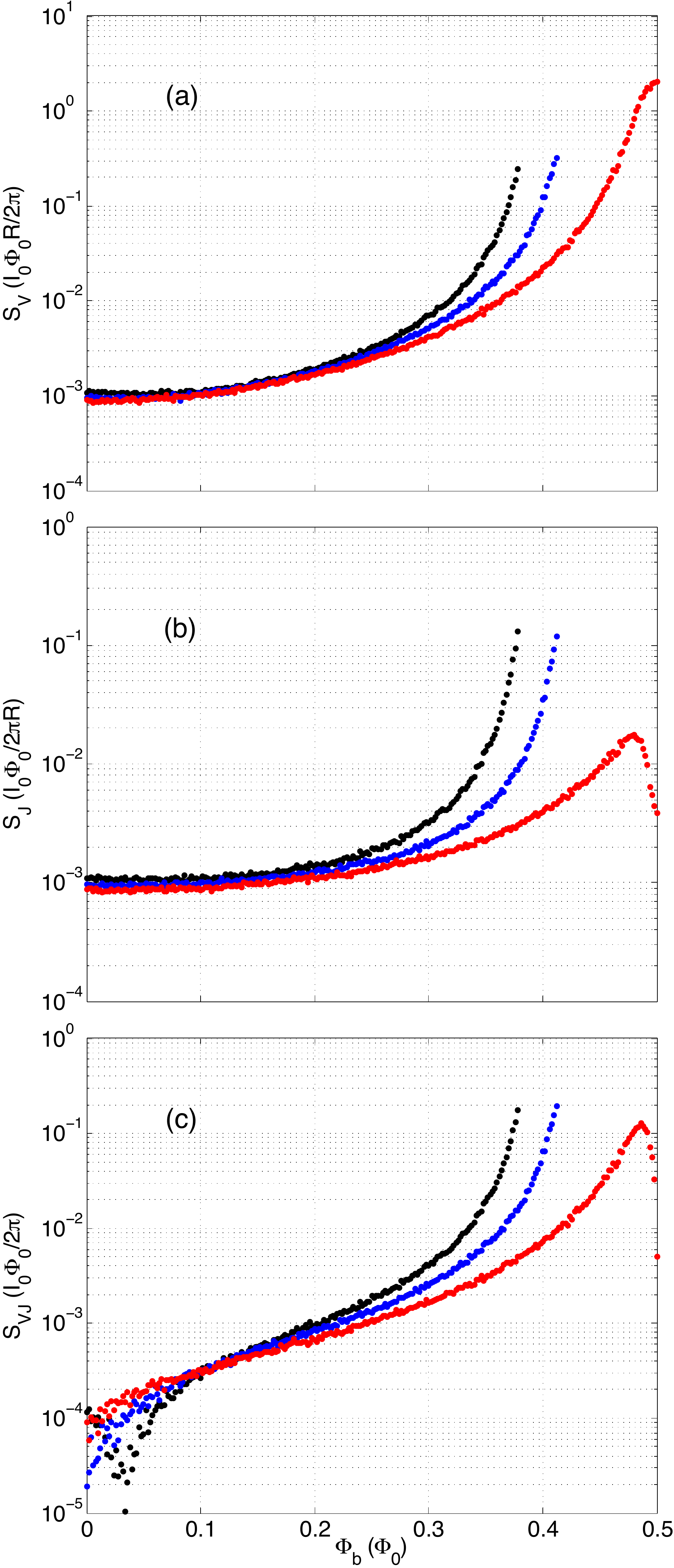}
\vspace*{-0.0in} \caption{Quantum noises (a) $S_V$, (b) $S_J$, and (c) $S_{VJ}$ \textit{versus} flux for bias currents $I_b = 1.8 \, I_0$ (black), $ 1.9 \, I_0$ (blue), and $2.0 \, I_0$ (red). The SLUG parameters are $\beta_L = 1$, $\beta_C = 0.8$, $L$~=~10~pH, and $C$~=~50~fF.}
\label{fig:quantumnoise}
\end{center}
\end{figure}
We have calculated the added noise of the symmetric SLUG in the zero-temperature limit, where the voltage spectral density of the shunt resistors becomes $2 h f R$. We generate a single-sided quantum spectral density by digitally filtering gaussian white noise. Using the quantum noise as a driving term in the Langevin equations \eqref{eqn:SLUG}, we evaluate the voltage power spectral density $S_V(f)$ at the device output, the circulating current spectral density $S_J(f)$, and the cross spectral density $S_{VJ}(f)$; in Fig. \ref{fig:quantumnoise} we plot these noises \textit{versus} flux for various bias currents. Once again, the device noise temperature $T_n$ can be evaluated from the circuit of Fig. \ref{fig:noise_analysis}. We assume a zero-temperature source impedance $Z_s = R_s + j X_s$, and equate the total noise of the amplifier to the noise contribution from a source resistance $R_s$ at a finite effective temperature $T_n$. The amplifier noise temperature is obtained from the relation
\begin{align}
&2 h f R_s \, \textrm{coth}\left(hf/2k_B T_n\right) \frac{V_\Phi^2 M^2}{R_t^2 + X_t^2} =
\nonumber \\
 &S_V + S_J \frac{\omega^2 V_\Phi^2 M^4}{R_t^2 + X_t^2} + 2 S_{VJ} \frac{\omega V_\Phi M^2 X_t}{R_t^2 + X_t^2} + 2 h f R_s \frac{V_\Phi^2 M^2}{R_t^2 + X_t^2}.
\end{align}
Alternatively, we can express the noise contribution of the device in terms of an added number of noise photons $n$, where $n$ and $T_n$ are related as follows:
\begin{align}
\textrm{coth}\left(hf/2k_B T_n\right) = 2 n + 1,
\label{eqn:n_Tn}
\end{align}
so that
\begin{align}
n = \frac{1}{2 h f R_s} \left[\frac{S_V}{2} \, \frac{R_t^2 + X_t^2}{V_\Phi^2 M^2} + \frac{S_J}{2} \omega^2 M^2 + S_{VJ} \frac{\omega}{V_\Phi} X_t \right].
\label{eqn:n}
\end{align}
The optimal source impedance $Z_{s,opt} = R_{S,opt} + j X_{s,opt}$ is obtained from the relations $\partial n / \partial X_t~=~0$ and $\partial n / \partial R_s~=~0$. The imaginary part of the optimal source impedance is given as follows:
\begin{align}
X_{s,opt} = -\frac{S_{VJ}}{S_V} \omega V_\Phi M^2 - X_i.
\end{align}
Similarly, the real part of the optimal source impedance is written
\begin{align}
R_{s,opt} = \left[1 + \left(\frac{V_\Phi R}{\rho_i \omega S_V}\right)^2 \left(S_V S_J - S_{VJ}^2 \right)\right]^{1/2} R_i.
\end{align}
In the limit $V_\Phi \gg \omega$, we find
\begin{align}
R_{s,opt} \approx \frac{\omega V_\Phi M^2}{S_V} \left(S_V S_J - S_{VJ}^2 \right)^{1/2}.
\label{eqn:rsquant}
\end{align}
In Fig. \ref{fig:Rsopt_quantum} we plot $R_{s,opt}/R_i$ in the quantum regime \textit{versus} flux for a range of bias currents. 

\begin{figure}[t]
\begin{center}
\includegraphics[width=.49\textwidth]{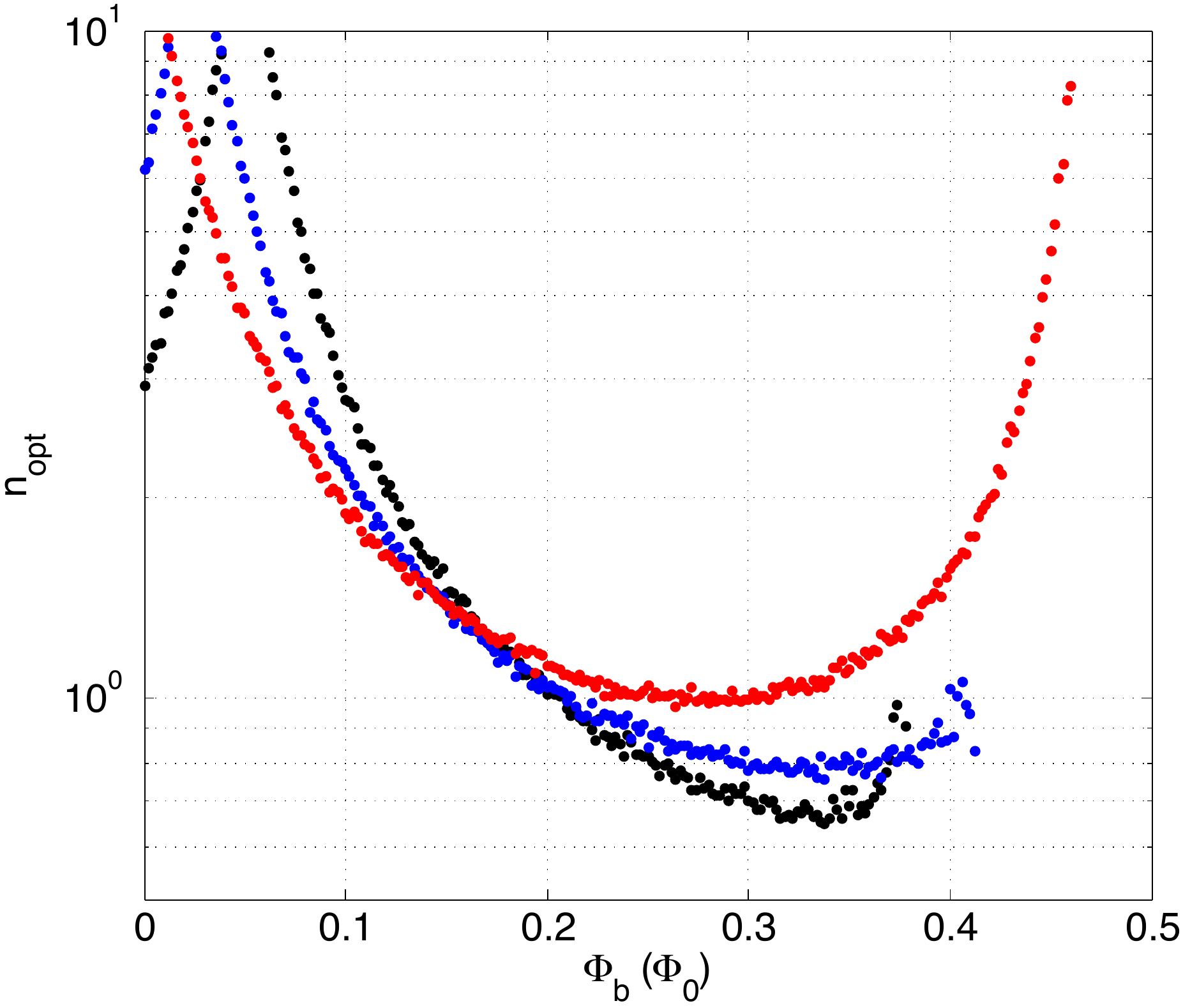}
\vspace*{-0.0in} \caption{Minimum number of added noise photons in the quantum regime $n_{opt}$ \textit{versus} flux for bias currents $I_b = 1.8 \, I_0$ (black), $ 1.9 \, I_0$ (blue), and $2.0 \, I_0$ (red). The operating frequency is 5~GHz and the SLUG parameters are $\beta_L = 1$, $\beta_C = 0.8$, $L$~=~10~pH, and $C$~=~50~fF.}
\label{fig:n_added}
\end{center}
\end{figure}

For the optimally matched source, the added number of noise photons is given by
\begin{align}
n_{opt} = \frac{1}{2 \hbar V_\Phi} \left(S_V S_J - S_{VJ}^2 \right)^{1/2}.
\end{align}
In Fig. \ref{fig:n_added} we plot $n_{opt}$ \textit{versus} flux, for various current biases. We see that for an appropriately noise-matched source the SLUG approaches a noise level that is close to the standard quantum limit $n_{SQL} = 1/2$, the minimum achievable added noise for a phase-insensitive linear amplifier \cite{Caves82}.

\section{VII. Amplifier Design}

\begin{figure}[t]
\begin{center}
\includegraphics[width=.49\textwidth]{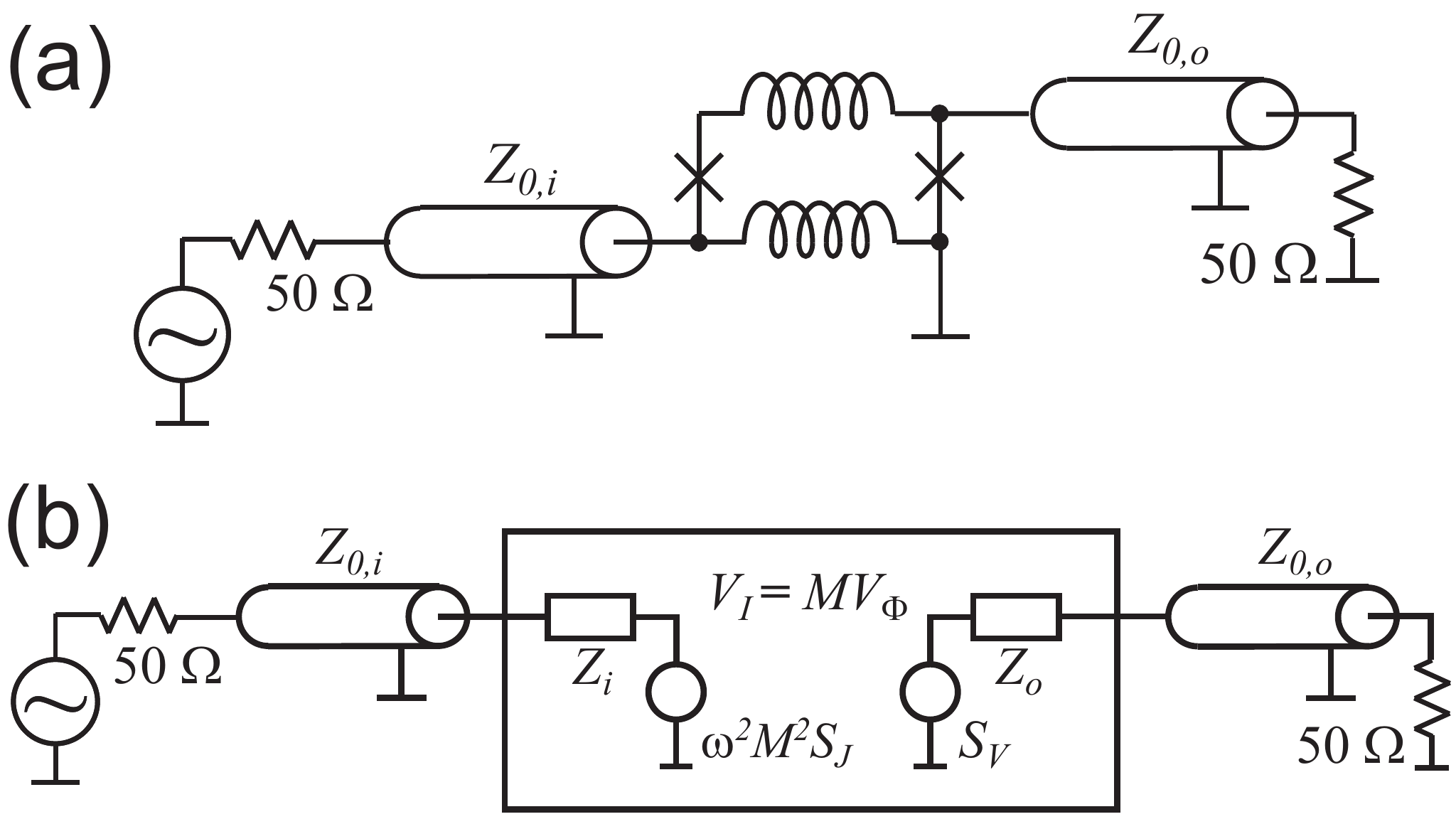}
\vspace*{-0.0in} \caption{(a) Schematic of SLUG microwave amplifier. (b) Circuit for amplifier analysis.}
\label{fig:amplifier}
\end{center}
\end{figure}

The above analysis demonstrates that the SLUG is an attractive gain element for the realization of a low-noise microwave amplifier. We now consider concrete external networks used to embed the device in a 50~$\Omega$ environment. The tasks are to maximize power transfer to and from the device and to match the 50~$\Omega$ source to the optimal noise impedance at the desired operating frequency. For example, to maximize gain we design a conjugate matching network to transform the 50~$\Omega$ source to $R_i - j X_i$. On the other hand, optimal noise performance is achieved for an input matching network that transforms the 50~$\Omega$ generator to the complex optimal source impedance $Z_{s,opt} = R_{s,opt} + j X_{s,opt}$. Since $R_{s,opt} \gg R_i$ for typical parameters, it is generally not possible to achieve a simultaneous power match and noise match. However, it is possible to find a compromise where there is reasonable gain and good noise performance over a relatively broad bias range. Fig. \ref{fig:amplifier}a shows a schematic diagram of a SLUG-based microwave amplifier with transmission line matching sections at the input and output. To calculate amplifier gain and noise performance, we treat the SLUG as a ``black box" with scattering and noise parameters derived from the calculations of Sections IV-VI (Fig. \ref{fig:amplifier}b).

\begin{figure}[t]
\begin{center}
\includegraphics[width=.49\textwidth]{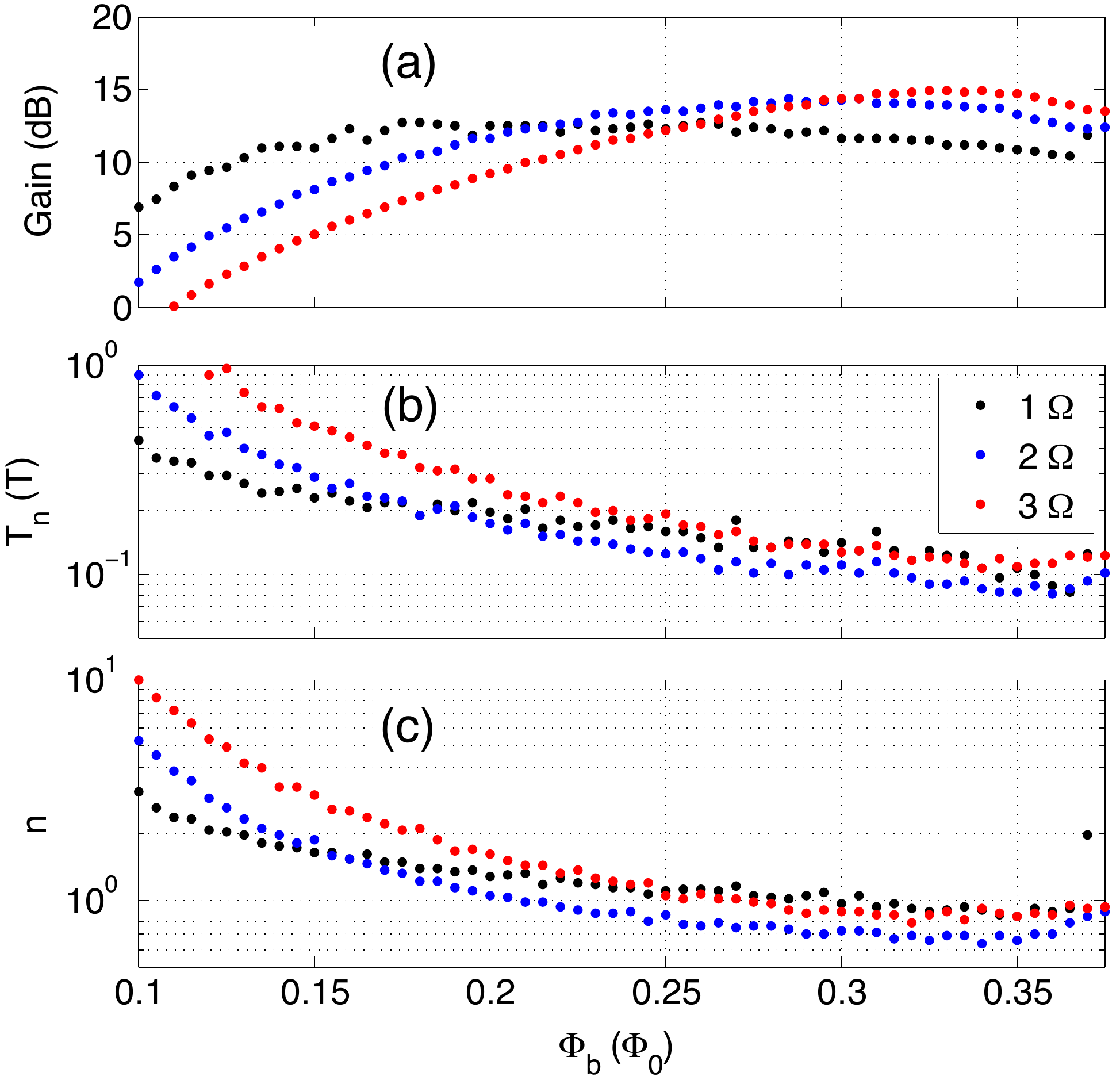}
\vspace*{-0.0in} \caption{(a) Gain, (b) noise temperature, and (c) added noise quanta for a 5~GHz amplifier incorporating a 10~pH SLUG element with $\beta_L = 1$, $\beta_C = 0.8$, $C$~=~50~fF and $I_b = 1.8 \, I_0$. The input matching network is a single transmission line section with characteristic impedance as indicated. Gain and added noise are evaluated at the frequency where the quantum noise contribution of the SLUG is minimum.}
\label{fig:quadplot}
\end{center}
\end{figure}

As an example, we show in Fig. \ref{fig:quadplot} the frequency-dependent gain, noise temperature $T_n$, and added noise quanta $n$ for SLUG amplifiers operated with different single-section transmission line input couplers with characteristic impedance in the range from 1-3 $\Omega$. Here we have used the full expressions \eqref{eqn:Tnthermal} and \eqref{eqn:n} to calculate the frequency-dependent noise contribution of the amplifier in the thermal and quantum regimes, respectively. The length of the input coupler provides a bare quarter-wave resonance at 6.5 GHz; inductive loading by the SLUG pulls the operating frequency down to the desired value of 5 GHz. 
We remark that the transmission line impedances considered here are readily achieved with thin-film microstrip technology: for example, a trace width of 10~$\mu$m and a dielectric with $\epsilon_r = 4$ and thickness 100~nm corresponds to a characteristic impedance of 2~$\Omega$.

\begin{figure}[t]
\begin{center}
\includegraphics[width=.49\textwidth]{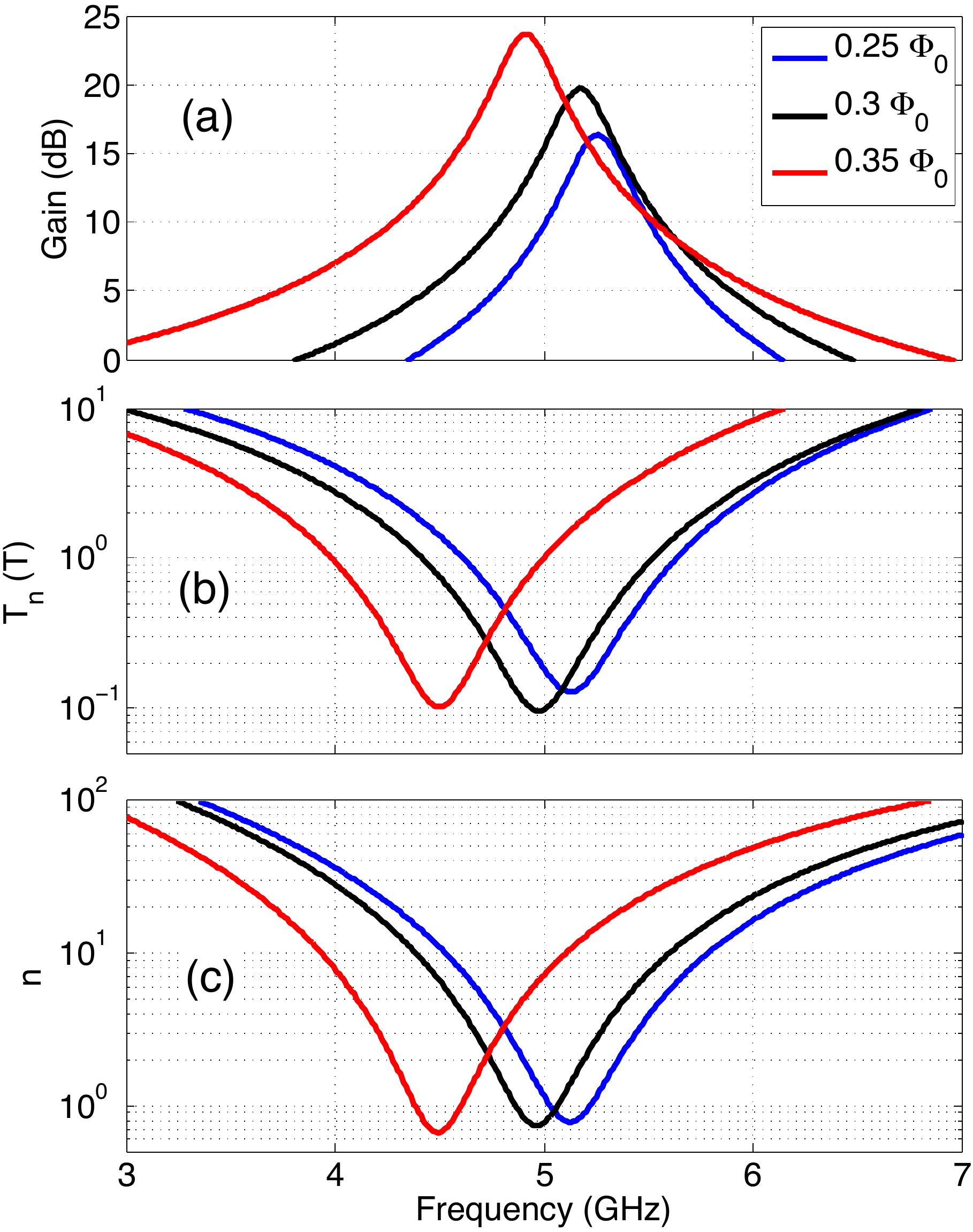}
\vspace*{-0.0in} \caption{(a) Gain, (b) noise temperature in the thermal regime, and (c) added noise in the quantum regime for a 5~GHz SLUG amplifier. The device parameters are $\beta_L = 1$, $\beta_C = 0.8$, $L$~=~10~pH, $C$~=~50~fF, and $I_b = 1.8 \, I_0$. The input matching network is a single transmission line section with bare quarter-wave resonance at 6.5~GHz and characteristic impedance 2~$\Omega$.}
\label{fig:singlesection}
\end{center}
\end{figure}

In Fig. \ref{fig:singlesection} we consider the frequency-dependent gain and noise performance of SLUG amplifiers operated with different fixed single-section input coupling networks. Due to the nonvanishing cross spectral density $S_{VJ}$, the minimum noise temperature occurs at a frequency that is somewhat lower than the frequency of maximum gain. For a $Z_{0,i}=2$~$\Omega$ input coupler, we achieve noise within 50\% of the standard quantum limit at a frequency where amplifier gain is 15~dB, and noise within a factor of 2 of the standard quantum limit at a frequency where gain is 18~dB.

\begin{figure}[t]
\begin{center}
\includegraphics[width=.49\textwidth]{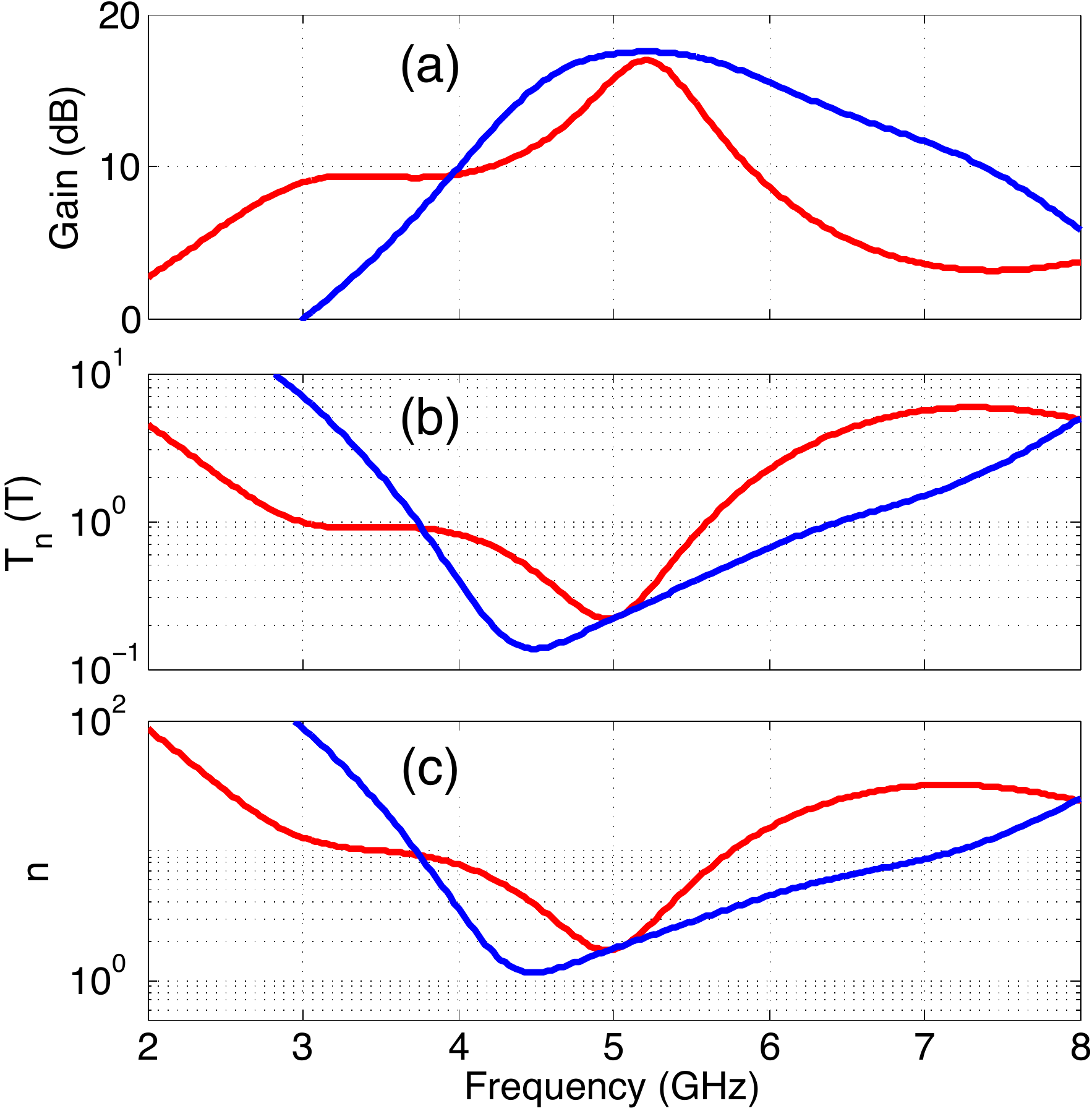}
\vspace*{-0.0in} \caption{(a) Gain, (b) noise temperature in the thermal regime, and (c) added noise in the quantum regime for broadband amplifiers incorporating a 10~pH SLUG element with $\beta_L = 1$, $\beta_C = 0.8$, $I_b = 1.8 \, I_0$, and $\Phi_b = 0.35 \, \Phi_0$. The red traces correspond to a three-section input matching network with quarter-wave resonances at 5 GHz and with characteristic impedances of 24.3 $\Omega$, 17.4 $\Omega$, and 3.0 $\Omega$, derived from numerical minimization of the SLUG quantum noise over the band from 4.5 GHz to 5.5 GHz. The blue traces correspond to a matching network consisting of three sections with characteristic impedance 29.8 $\Omega$, 7.1 $\Omega$, and 1.1 $\Omega$ followed by a series capacitance of 38 pF to tune out the imaginary part of the SLUG input impedance at a frequency of 5 GHz.}
\label{fig:three_section}
\end{center}
\end{figure}

Finally, we note that is possible to increase amplifier bandwidth significantly by coupling the input signal to the device via a multisection transformer with stepped characteristic impedances. As an example, we show in Fig. \ref{fig:three_section} the frequency-dependent gain and added noise for amplifiers operated with different three-section matching networks. Here the length of each transmission line section is chosen to provide a bare quarter-wave resonance at 5 GHz, and the characteristic impedances were determined by numerical minimization of the quantum noise contribution of the SLUG in the frequency range from 4.5 to 5.5 GHz. 

\section{VIII. Dynamic Range}

\begin{figure}[t]
\begin{center}
\includegraphics[width=.49\textwidth]{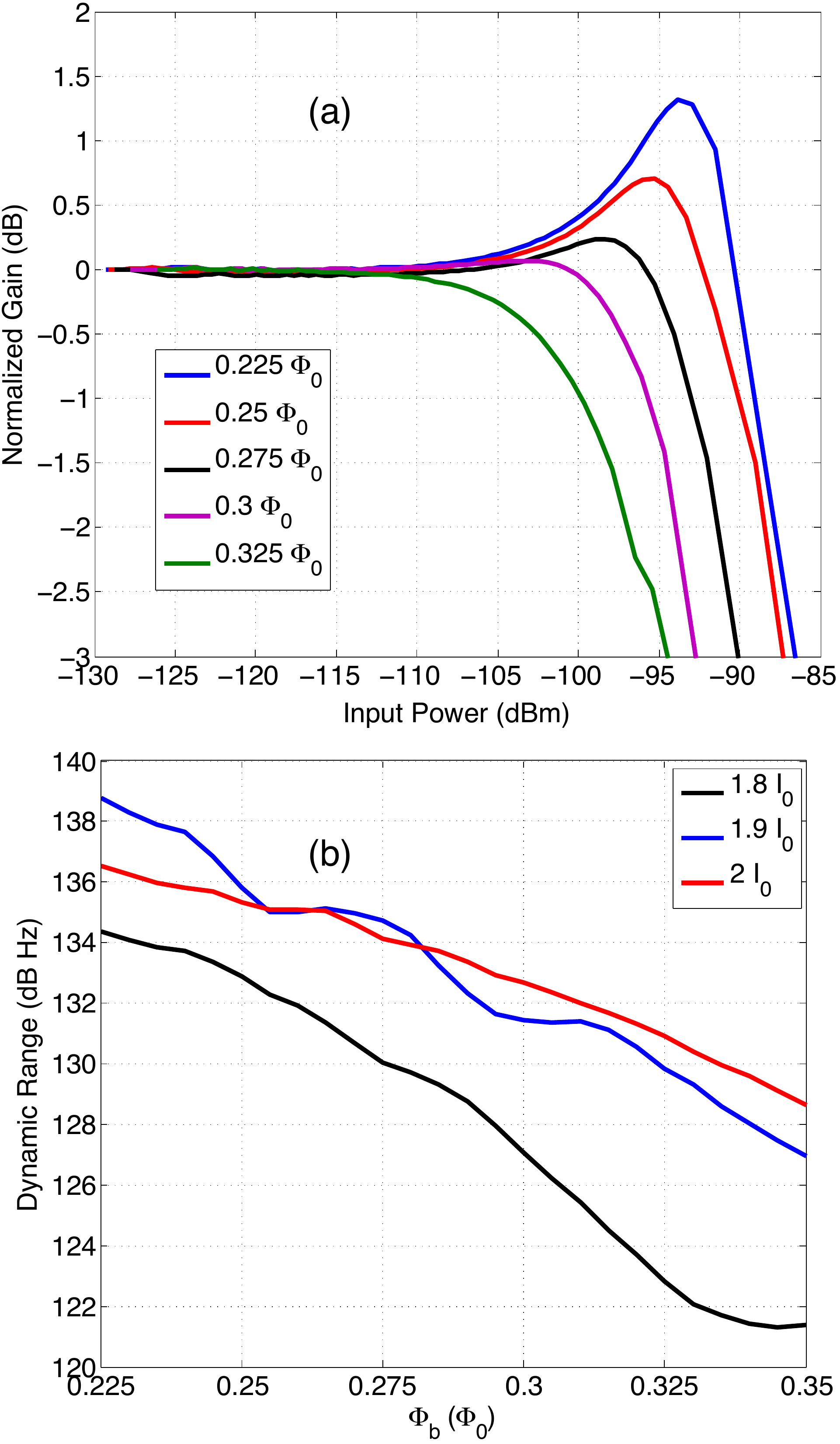}
\vspace*{-0.0in} \caption{(a) Normalized gain \textit{verus} input power for a SLUG element with $\beta_L = 1$, $\beta_C = 0.8$, $L= 10$~pH, $C=50$~fF, and $I_b = 1.8 \, I_0$. The different traces correspond to various flux bias points. (b) SLUG dynamic range \textit{versus} flux for various current bias points; the device parameters are as in (a), and we assume a zero-temperature quantum spectral density for the SLUG shunt resistors.}
\label{fig:dynamic_range}
\end{center}
\end{figure}

The strong nonlinearity of the SLUG leads to gain compression and harmonic generation when the device is driven with a large-amplitude signal. It is important to verify that the SLUG dynamic range will be sufficient for the desired application. In Fig. \ref{fig:dynamic_range}a we plot normalized SLUG gain \textit{versus} signal power coupled to the device input over a range of bias parameters for $\beta_L = 1$, $\beta_C = 0.8$, $L$~=~10~pH and $C$~=~50~fF. These plots were generated by solving the SLUG equations of motion (\ref{eqn:SLUG}) with a sinusoidal driving term of varying amplitude. Depending on bias point, the 1 dB compression point occurs somewhere in the range from -110~dBm to -~90~dBm, corresponding to input powers from 10~fW to 1~pW. These 1~dB compression points are comparable to those seen in other SQUID-based microwave amplifiers \cite{Spietz09} and 1-2 orders of magnitude higher than those achieved with typical Josephson parametric amplifiers \cite{Yamamoto08}. Amplifier dynamic range is determined by dividing the signal power at 1 dB compression by the noise power contributed by the SLUG over a given bandwidth. In Fig. \ref{fig:dynamic_range}b we plot SLUG dynamic range; here we have used the zero-temperature quantum spectral density for the shunt resistors of the SLUG. We find a typical value of 130~dB~Hz, corresponding to a dynamic range of 40~dB in an amplifier bandwidth of 1~GHz. For applications related to dispersive readout of qubits in a circuit QED architecture, where the focus is on measurement of signals at the level of single microwave quanta in bandwidths of order 100~MHz to 1~GHz, the dynamic range of the SLUG amplifier is more than adequate.

\section{IX. Effect of Input Circuit Admittance}

In the above analysis, we have solved for the behavior of the isolated SLUG element, and then treated the device as a ``black box" with known scattering parameters for the purpose of designing appropriate matching networks. In reality, the nonvanishing admittance at the device input and output will modify the device characteristics, and a complete treatment must take loading by the external circuit into account. The scattering parameters will now depend on the particular realization of the matching network, and a full exploration of the space of design parameters becomes tedious. However, we find that the performance of the SLUG amplifier is not greatly affected by the nonvanishing input circuit admittance, particularly once modest steps are taken to decouple the SLUG element from the higher-order modes of the resonant input matching network.

\begin{figure}[t]
\begin{center}
\includegraphics[width=.49\textwidth]{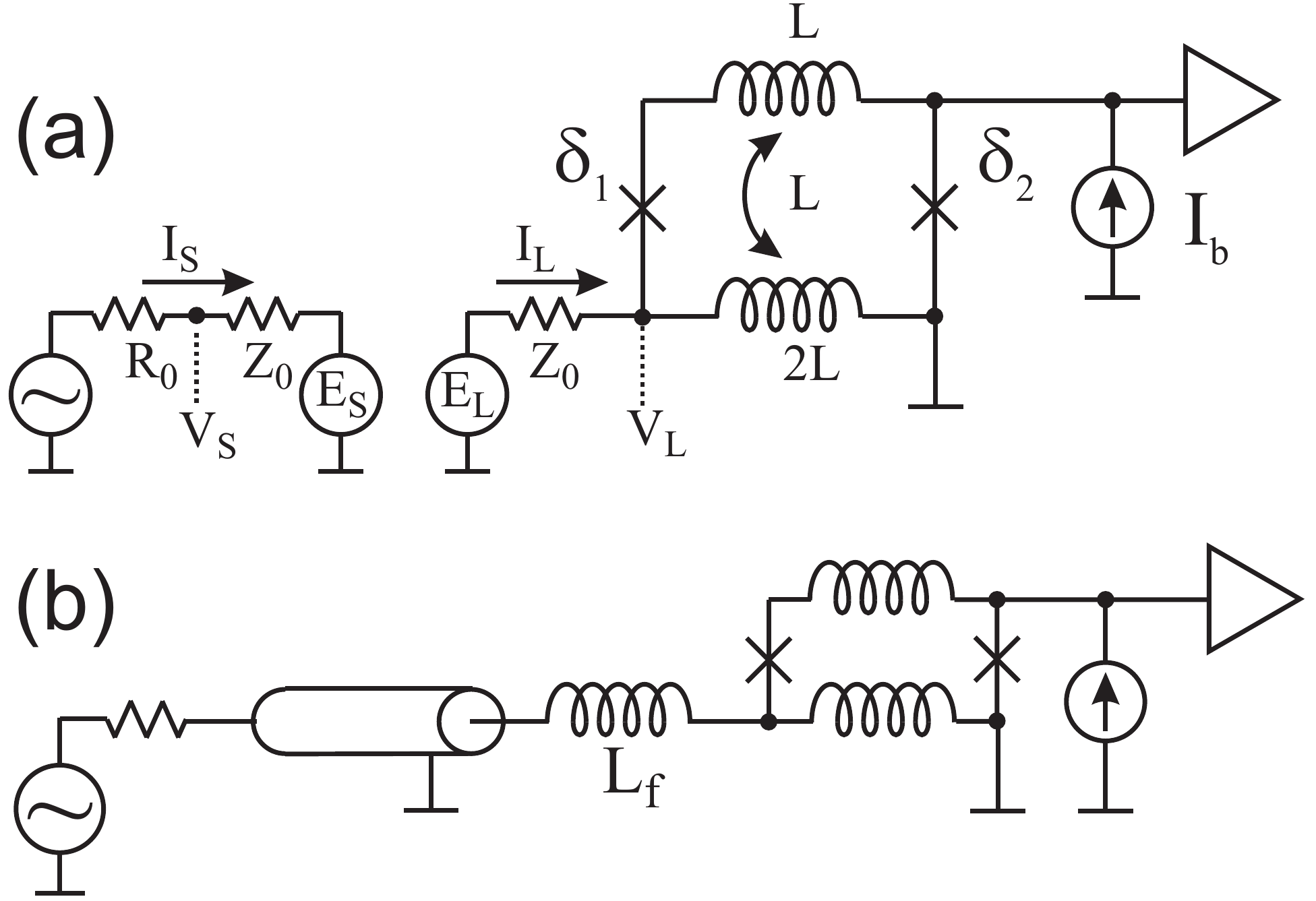}
\vspace*{-0.0in} \caption{(a) Model for circuit analysis with finite input circuit admittance. (b) Amplifier circuit with filter inductor $L_f$ to decouple SLUG from modes of the input circuit.}
\label{fig:filter_circuits}
\end{center}
\end{figure}
\begin{figure}[t]
\begin{center}
\includegraphics[width=.49\textwidth]{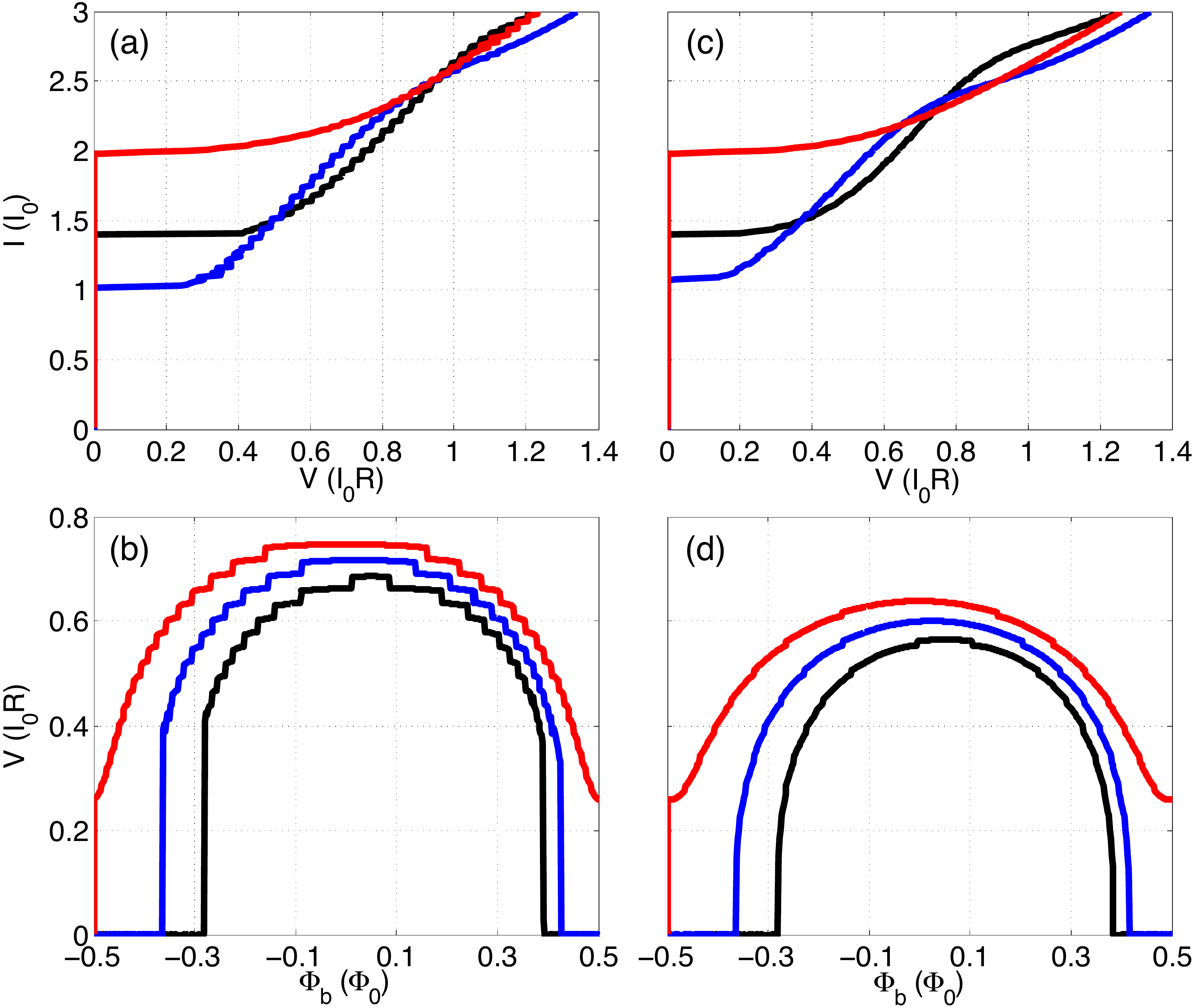}
\vspace*{-0.0in} \caption{(a) I-V curves of a SLUG operated with a transmission line input circuit with characteristic impedance $Z_0$~=~2~$\Omega$ and bare quarter-wave resonance at 10~GHz for $\Phi_b = 0 \, \Phi_0$ (black), $ 0.25 \, \Phi_0$ (blue), and $0.5 \, \Phi_0$ (red). (b) V-$\Phi$ curves of the same SLUG for $I_b = 1.8 \, I_0$ (black), $ 1.9 \, I_0$ (blue), and $2.0 \, I_0$ (red). (c)-(d) As in (a)-(b), respectively, for a circuit incorporating a 60~pH filter inductor $L_f$ to decouple the modes of the SLUG from the modes of the input circuit.}
\label{fig:filter_dc}
\end{center}
\end{figure}

To take into account the admittance of the resonant input matching network, we modify the junction equations of motion \eqref{eqn:SLUG} to include an additional term representing the current drawn by the input circuit. The circuit model is shown in Fig. \ref{fig:filter_circuits}a. The input transmission line of impedance $Z_0$ can be exactly modeled as a pair of coupled, time dependent voltage sources $E_L$ and $E_S$. These are related to the voltages $V_{L,S}$ and currents $I_{L,S}$ at the two ends of the transmission line as follows:
\begin{align}
E_L(t) &= V_S(t-t_D) + Z_0 I_S(t - t_D) 
\nonumber \\
E_S(t) &= V_L(t-t_D) - Z_0 I_L(t - t_D),
\end{align}
where $t_D$ is the propagation delay along the transmission line. The input current is then determined by the additional differential equation
\begin{equation}
\dot{I_L} = \frac{1}{L} \left[\frac{\Phi_0}{2 \pi} (\dot{\delta_2} - \dot{\delta_1}) - E_L + I_L Z_0 \right].
\end{equation}
Using the modified equations of motion for the junction phases, we calculate the dc characteristics of the SLUG. The I-V and V-$\Phi$ curves of a 10~pH, $\beta_L = 1$ SLUG with a 10~GHz quarter-wave input transformer are shown in Fig. \ref{fig:filter_dc}a-b. We observe sharp Shapiro step-like structure at voltages corresponding to Josephson frequencies that are integer multiples of the half-wave resonance of the input circuit. While quantum fluctuations of the SLUG shunts smooth out this structure somewhat, it is clearly desirable to decouple to the SLUG from the higher-order standing wave modes of the input circuit, as these modes will limit amplifier dynamic range and lead to excess noise.

To suppress the resonances of the input circuit, we insert a filter inductor $L_f$ of order tens of pH between the input coupler and the SLUG element, as shown in Fig. \ref{fig:filter_circuits}b. In Fig. \ref{fig:filter_dc}c-d we plot the SLUG characteristics with a 60~pH filter inductor in place. We see that the resonant structure is greatly suppressed.

\begin{figure}[t]
\begin{center}
\includegraphics[width=.49\textwidth]{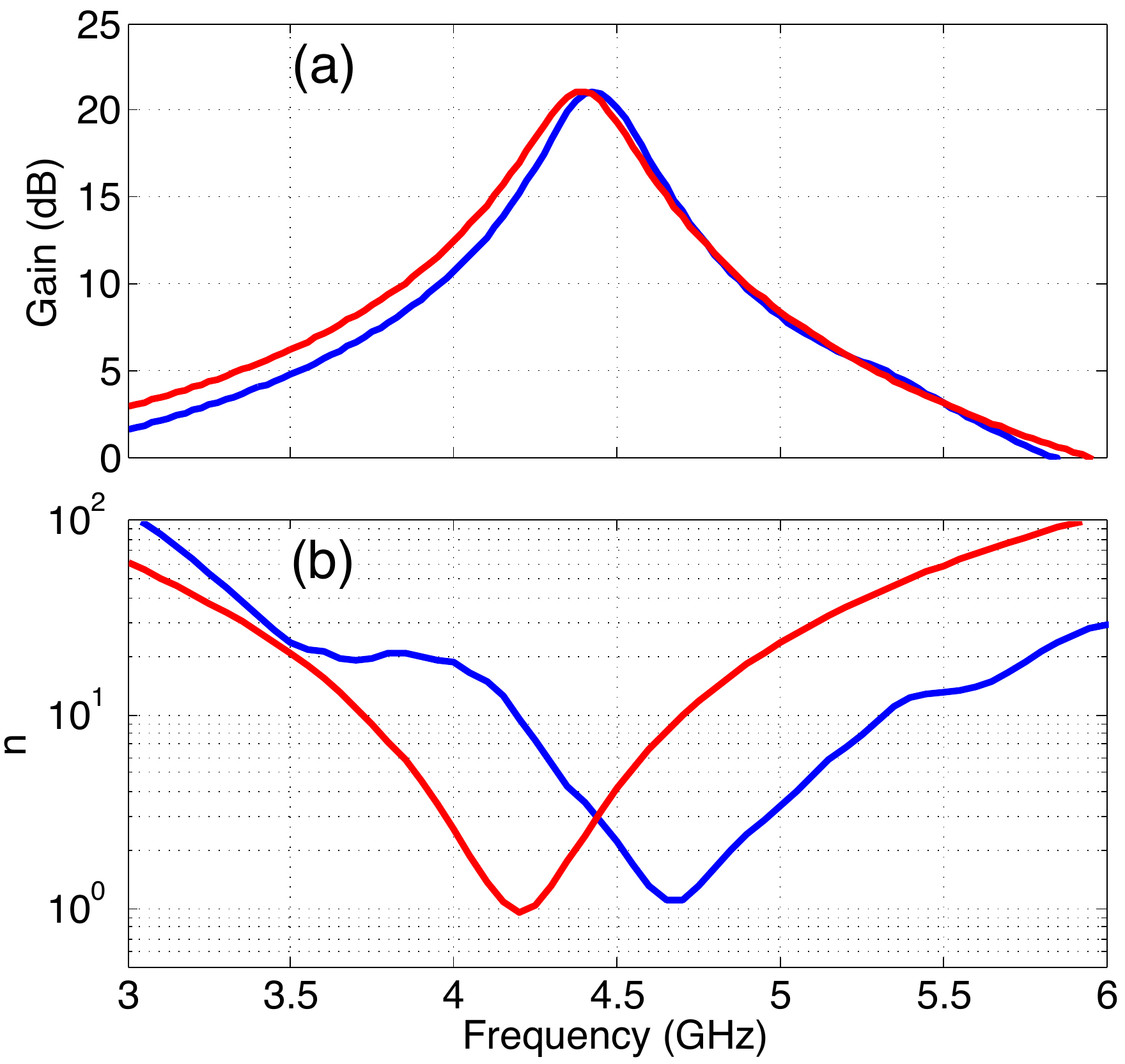}
\vspace*{-0.0in} \caption{(a) Gain and (b) added noise in the quantum regime for SLUG amplifiers calculated using the ``black box" scattering parameters of the isolated SLUG (red traces) or by solving the full circuit model of Fig. \ref{fig:filter_circuits}. The SLUG parameters are $\beta_L = 1$, $\beta_C = 0.8$, $L= 10$~pH, $C=50$~fF, $I_b = 1.8 \, I_0$, and $\Phi_b = 0.35 \Phi_0$. The input matching network consists of a 2 $\Omega$ transmission line section with bare quarter-wave resonance at 10 GHz followed by a filter inductor $L_f = 60$ pH.}
\label{fig:full_circuit}
\end{center}
\end{figure}

We can now calculate the gain and noise properties of the complete circuit of Fig. \ref{fig:filter_circuits}b by performing a full integration of the amplifier equations of motion. Power gain and bandwidth are determined by driving the amplifier with a sinusoidal input tone and monitoring the SLUG output at the excitation frequency. In Fig. \ref{fig:full_circuit}a we plot frequency-dependent gain for the SLUG circuit. The blue trace is the result of the full circuit simulation, where we have taken a transmission line input with characteristic impedance $Z_0 = 2$ and a length corresponding to a bare quarter-wave resonance at 10~GHz, significantly higher than the amplifier operating frequency of 4.5~GHz in order to compensate for the additional reactive loading by the filter inductor. The red trace was obtained by treating the SLUG as a ``black box" with scattering parameters calculated as described above in Section IV. The agreement with the full circuit simulation is good, confirming that the filter inductance has effectively isolated the modes of the SLUG and the input circuit.

To calculate the frequency-dependent noise temperature $T_n(f)$, we simulate a ``hot load / cold load" experiment where we compare the power spectra $S_{V,cold}$ and $S_{V,hot}$ at the device output for source resistances at temperatures $T=0$ and $T_b$, respectively. In the thermal regime, we have
\begin{align}
T_n(f) = \frac{S_{V,cold}(f)}{S_{V,hot}(f) - S_{V,cold}(f)} \,\, T_b.
\end{align}
In the quantum regime, we find
\begin{align}
\frac{\textrm{coth}\left[hf/2k_B(T_b+T_n) \right]}{\textrm{coth}\left( hf/2k_B T_n \right)} = \frac{S_{V,hot}}{S_{V,cold}}.
\end{align}
The added noise number is then obtained from Eq. \ref{eqn:n_Tn}.
In Fig. \ref{fig:full_circuit}b we plot the added noise of a 5~GHz SLUG amplifier calculated with the full circuit model and with the ``black box" scattering parameters of the isolated SLUG. The noise magnitude is similar in the two cases, although the full circuit solution predicts a higher frequency for the minimum in the amplifier noise contribution. We understand the shift in the frequency-dependent noise characteristics to be due to a modification of the circulating current spectral density $S_J$ by the nonvanishing admittance of the input network.

\section{X. Hot Electron Effects}

At millikelvin temperatures electrons decouple from the phonons, and the electron temperature of the SLUG shunts can be significantly higher than the bath temperature. Wellstood \textit{et al.} showed that the electron temperature $T_e$ in a metal thin film resistor is given by
\begin{align}
T_e = (P/\Sigma \Omega + T_p^5)^{1/5},
\end{align}
where $P$ is the power deposited in the resistor, $\Sigma$ is a materials parameter equal to approximately $2~\times~10^9$~W/m$^3$K$^5$, $\Omega$ is the normal metal volume, and $T_p$ is the phonon temperature \cite{Wellstood94}. The elevated temperature of the shunt resistors translates directly to elevated noise temperature of the amplifier. For a device with fixed $\beta_C$, the power dissipation in the shunts scales as $1/R^3$. Hot electron effects will be particularly relevant for the microwave amplifiers discussed here, as optimal performance is achieved for small SLUG inductance, corresponding to large critical currents and small shunt resistances.

A proven strategy to promote thermalization of the SLUG shunts at millikelvin temperatures is to fabricate large-volume normal metal cooling fins in metallic contact with the resistor element. At low temperatures, the inelastic diffusion length is of order several mm \cite{Wellstood94}; the cooling fins thus allow hot electrons generated in a localized region of the shunt resistor to diffuse over a large volume and thermalize with cold electrons and phonons. Wellstood \textit{et al.} demonstrated a significant reduction in the electron temperature of dc SQUIDs incorporating 400~$\times$~400~$\mu$m$^2$ CuAu cooling fins with thickness around 1~$\mu$m, with measured electron temperatures under 40~mK \cite{Wellstood89}. A similar approach has been used to suppress hot-electron effects and reduce the noise temperature of microstrip SQUID amplifiers operated in the radiofrequency regime \cite{Kinion11}. It will be straightforward to integrate normal metal cooling fins with area of order 1~mm$^2$ into a standard microwave SLUG amplifier geometry without compromising the microwave integrity of the circuit. We anticipate that the addition of such cooling fins will make it possible to attain electron temperatures under 100~mK for the device parameters considered here, corresponding to operation far in the quantum regime for frequencies in the range from 5-10~GHz.

\section{XI. Concluding Remarks}

We have presented a comprehensive theoretical treatment of the SLUG microwave amplifier. Specific advantages of this approach over competing approaches to low-noise microwave amplification are as follows:
\begin{enumerate}
\item{The low-inductance device geometry is compact, straightforward to model at microwave frequencies, and readily integrated into a microwave transmission line.}
\item{The device input and output are both reasonably well-matched to a 50~$\Omega$ transmission-line impedance, facilitating broadband operation. Moreover, multisection transmission-line input couplers provide a clear path to attaining bandwidths of order GHz while maintaining excellent gain and noise performance.}
\item{It is straightforward to decouple the SLUG modes from the input modes, allowing separate optimization of the gain element and the input matching network.}
\item{The dynamic range of the amplifier is large relative to that required for qubit readout or circuit QED applications.}
\item{Due to its extremely small magnetic sensing area, the SLUG gain element is extremely robust and immune to ambient magnetic field fluctuations.}
\end{enumerate}

We believe that we have identified the major technical obstacles and outlined a clear path to device optimization. We anticipate that these amplifiers will be attractive in the context of qubit readout in a circuit QED architecture \cite{Wallraff04}, either as a near quantum-limited first-stage amplifier or as an ultralow noise postamplifier following a Josephson paramp. Other possible applications include fundamental studies of microwave photon counting statistics \cite{Bozyigit11}, or ultralow noise amplification for dark-matter axion detection \cite{Asztalos10}.

\begin{acknowledgments}
We thank J.M. Martinis, M. M$\ddot{\textrm{u}}$ck, and B.L.T. Plourde for useful discussions, and we acknowledge support from IARPA under contracts W911NF-10-1-0324 and W911NF-10-1-0334. All statements of fact, opinion or conclusions contained herein are those of the authors and should not be construed as representing the official views or policies of the U.S. Government.
\end{acknowledgments}

\bibliography{master_refs}

\end{document}